\documentclass[onecolumn]{aastex62}
\pdfoutput=1
\usepackage{amsmath} 
\usepackage{natbib}
\usepackage{mathtools} 
\usepackage{mathrsfs}
\usepackage{hyperref}

\shorttitle{Cylindrical Poisson Solver}
\shortauthors{Moon, Kim, \& Ostriker}

\begin{document}

\title{A FAST POISSON SOLVER OF SECOND-ORDER ACCURACY FOR ISOLATED SYSTEMS IN THREE-DIMENSIONAL
CARTESIAN AND CYLINDRICAL COORDINATES}

\author[0000-0002-6302-0485]{Sanghyuk Moon}
\affiliation{Department of Physics \& Astronomy, Seoul National University, Seoul 08826, Korea}
\author[0000-0003-4625-229X]{Woong-Tae Kim}
\affiliation{Department of Physics \& Astronomy, Seoul National University, Seoul 08826, Korea}
\affiliation{Department of Astrophysical Sciences, Princeton University, Princeton, NJ 08544, USA}
\author[0000-0002-0509-9113]{Eve C. Ostriker}
\affiliation{Department of Astrophysical Sciences, Princeton University, Princeton, NJ 08544, USA}
\email{moon@astro.snu.ac.kr, wkim@astro.snu.ac.kr}
\email{eco@astro.princeton.edu}

\begin{abstract}
We present an accurate and efficient method to calculate the gravitational potential of an isolated system in three-dimensional Cartesian and cylindrical coordinates subject to vacuum (open) boundary conditions. Our method consists of two parts: an interior solver and a boundary solver. The
interior solver adopts an eigenfunction expansion method together with a tridiagonal matrix solver to solve the Poisson equation subject to the zero boundary condition.  The boundary solver employs James's method to calculate the boundary potential due to the screening charges required to keep the zero boundary condition for the interior solver. A full computation of gravitational potential
requires running the interior solver twice and the boundary solver once. We develop a method to compute the discrete Green's function in cylindrical coordinates, which is an integral part of the James algorithm to maintain second-order accuracy. We implement our method in the {\tt Athena++} magnetohydrodynamics code, and perform various tests to check that our solver is second-order accurate and exhibits good parallel performance.
\end{abstract}

\section{INTRODUCTION}

There are a number of astronomical systems, such as
galactic
and protostellar disks, where self-gravity and rotation play an essential role in dynamical evolution. For instance, starburst activity occurring in massive circumnuclear disks can not only inflate the natal disks to form thick tori surrounding active galactic nuclei (AGN) \citep{wada02,wada09} but also drive large-scale galactic winds and outflows \citep{strick00,strick04a,strick04b,sch18a,sch18b}. Accretion disks around AGN may be gravitationally unstable at some radii to form stars \citep{goodman03,goodman04,levin07,nayakshin07,jiang11}.
Self-gravity is also important in formation of large-scale spiral structure
\citep{goldreich65,baba13,onghia13} and giant molecular clouds \citep{kim03,dobbs08,tasker09} on larger scales in
galactic disks.
In addition, recent observations of young stellar objects indicate that at least in the early stage of evolution, protostellar disks are massive enough to be self-gravitating \citep{kratter16,tobin16}. Gravitational instability of such disks may form trailing spirals that can redistribute the mass and angular momentum and induce heat via shocks \citep{mejia05, evans15} and may be responsible for the formation of giant planets \citep{boss07,zhu12}.

To follow evolution of self-gravitating
disks, one needs to solve the Poisson equation
\begin{equation}\label{eq:Poisson}
  \nabla^2\Phi = 4\pi G \rho,
\end{equation}
in cylindrical coordinates $(R,\phi,z)$ subject to a proper boundary condition. In Equation \eqref{eq:Poisson}, $\Phi$, $\rho$, and $G$ refer to the gravitational potential, mass density, and gravitational constant, respectively. For an isolated system, $\Phi$ has to satisfy
vacuum (or ``open'') boundary conditions
(i.e., $\Phi$ vanishes at infinite distances), for which the formal solution of Equation \eqref{eq:Poisson} is given by
\begin{equation}\label{eq:intPoisson}
  \Phi({\bf x}) = \iiint {\cal G}_\infty({\bf x,x'}) \rho({\bf x'}) \,d^3x',
\end{equation}
where ${\cal G}_\infty({\bf x,x'})  \equiv - G/ |{\bf x-x'}|$ is the gravitational potential per unit mass due to a point source situated at ${\bf x'}$. Hereafter, we call ${\cal G}_\infty({\bf x,x'})$ the continuous Green's function (CGF) to distinguish it from the discrete Green's function (DGF) based on the \emph{discrete} Laplace operator (e.g., \citealt{burk97}) discussed in Section \ref{s:dgf}.

In simulating dynamics of geometrically thin disks,
it has been customary to assume that the disk density along the vertical direction follows a simple function  such as Dirac's delta function for a razor-thin disk and a Gaussian function for a slightly extended disk
(e.g., \citealt{kal71,miller76, li09, wang15}).
In this case, the integral along the $z$-direction in Equation \eqref{eq:intPoisson}
can be performed  analytically, and finding $\Phi(R, \phi)$ at $z=0$ reduces to numerical evaluation of the remaining two-dimensional (2D) integral in the $R$--$\phi$ plane. For example, \citet{miller76} solved the gravitational potential of an infinitesimally-thin disk by using a fast Fourier transform (FFT) technique along the azimuthal direction, while directly summing the individual contributions from concentric rings. He introduced a constant softening factor in order to avoid singularity at $\bf x=x'$ of the CGF.
\citet{li09} applied this method to develop an efficient gravity solver for disks with finite thickness on a 2D \emph{uniform} polar grid. They improved the parallel efficiency of their  method by cutting off the high azimuthal Fourier modes based on the energy criterion.

When the grid spacing is \emph{logarithmic} in the radial direction, a suitable change of variables recasts the integral in Equation \eqref{eq:intPoisson} to a 2D convolution \citep{kal71,bt}, for which
the standard FFT convolution method works
efficiently \citep{hoc88}. For example, \citet{bm08} applied this technique to a razor-thin disk by taking a softening factor proportional to $R$ to avoid divergence of the CGF.
\citet{so08} extended this method to a slightly vertically-extended disk, in which finite disk thickness naturally provides the required softening.
Noting that softening reduces the accuracy of a gravity solver,
\citet{wang15} avoided singularity by using the force kernels integrated over cells, and achieved a second-order accuracy for self-gravity of a razor-thin disk.

Although the methods described above are useful and efficient, they are all limited to 2D polar geometry in the $R$--$\phi$ plane. To our knowledge, there is no efficient method available for fully three-dimensional (3D) cylindrical systems with the vacuum (open) boundary conditions. This is presumably because the Green's function integral takes a convolution form only along the azimuthal and vertical directions: there is no variable transformation that can cast the integral to a full 3D convolution. One may still attempt to perform the radial integral in Equation \eqref{eq:intPoisson} by direct summation, while applying the FFT convolution along the azimuthal and vertical directions. But, the associated computational cost is of order ${\cal O}(N^4+N^3\log N)$,
with $N$ being the typical number of cells in one spatial dimension
\citep{pfen93,sell97}, making the method computationally prohibitive.

In many cases, it is computationally more efficient to solve Equation \eqref{eq:Poisson} directly, rather than evaluating the integral in Equation \eqref{eq:intPoisson}. For example, \citet{gupta97} discretized Equation \eqref{eq:Poisson} using a fourth-order scheme in 2D Cartesian coordinates and employed a V-cycle multigrid method to solve the resulting linear system. \citet{lai07} adopted another fourth-order formula to discretize Equation \eqref{eq:Poisson} in cylindrical coordinates and solved the resulting linear system using FFT combined with a varient of the Bi-Conjugate Gradient iterative method. Perhaps, the most efficient and robust method to solve a discretized Poisson equation may be a full multigrid algorithm \citep[e.g.,][]{mat03}, which can in principle be implemented in either Cartesian or cylindrical coordinates.

However, all the methods mentioned above in turn require provision of
appropriate potentials at the domain boundaries in advance.
Since Equation \eqref{eq:intPoisson} naturally satisfies vacuum
boundary conditions, it is reasonable to use it to find
the desired boundary potentials for Equation \eqref{eq:Poisson}.
Still, the computational cost of ${\cal O}(N^4+N^3\log N)$ would be inevitable if radial summation is employed for the boundary potentials. One way to reduce the computational cost is to expand the Green's function in eigenfunction series and truncate it at some point. For instance, the so-called ``multipole expansion method'' \citep{bb75,zeus,boley08,katz16} in spherical polar coordinates costs ${\cal O}(l_{\rm max}m_{\rm max}N^3)$ operations for the boundary potential calculations, where $l_{\rm max}$  and $m_{\rm max}$ refer to the maximum  meridional and azimuthal mode numbers, respectively.  Although this method appears feasible for small $l_{\rm max}$ and $m_{\rm max}$,  the computational cost would increase to ${\cal O}(l_{\rm max}m_{\rm max}N^4)$ for a flattened mass distribution with surfaces lying close to domain boundaries. An additional $N$ factor in the computational cost arises from the fact that the interior and exterior multipole moments for such a flattened mass distribution are different at most boundary points (see, e.g., \citealt{cohl99}).

\citet{cohl99} derived an alternative expansion of the Green's function in cylindrical coordinates, which
they termed the compact cylindrical Green's function (CCGF). Their CCGF method can perfectly resolve a highly flattened mass distribution,
effectively using $l_{\rm max}=\infty$. Coupled with FFT, the CCGF method requires ${\cal O}(m_{\rm max}N^3 + N^3\log N)$ operations for the boundary potential calculations, and has been applied to 3D simulations of self-gravitating disks \citep[e.g.,][]{from05, mel08, mar12, motl17}.
When the mass distribution is highly non-axisymmetric and/or dominated by small-scale modes, however, the simulation outcomes depend rather sensitively on the choice of $m_{\rm max}$ \citep{from04b,from05}. In such situations,
an accurate force evaluation requires
$m_{\rm max}$ comparable to $N$ \citep{from05},
and the associated computational cost of ${\cal O}(m_{\rm max}N^3)$
as well as the memory requirement to store four-dimensional arrays
would be overwhelming for high-resolution simulations.

A very powerful method to deal with vacuum boundary conditions is the four-step
algorithm developed by \citet{james77}
\citep[see also][for more compact description]{mag07}.
In the first step, a preliminary solution $\Psi$  is computed based on the
interior density distribution and subject to the $\Psi=0$ boundary condition.
In the second step, $\Psi$ is used to compute a screening surface
charge, $\sigma  = \Delta^2 \Psi/(4 \pi G)$, and in the third step
a boundary potential
$\Theta$ generated by the surface charge and consistent with vacuum
boundary conditions is computed via convolution with the DGF.  In the
fourth step, the final  interior solution $\Phi$ is computed, making use of
$\Theta$ to enforce a new (nonzero) Dirichlet boundary condition.

The key physical principle underlying James's method is best
understood via an electrostatic analogy.  Consider the potential $\Phi$
produced by an isolated, charged box with a metal surface.  If the box is then
grounded, charges would flow in and be distributed over the surface to
enforce zero potential.  The potential $\Psi$ of the grounded box is
then the sum $\Psi = \Phi + \Theta$ of the potential
produced by the original internal charge
distribution plus the potential generated by the surface
charges.  On the surface of the box, the desired potential
produced by the original interior charge
distribution is then given by $\Phi^{\rm B}=-\Theta$, since
by definition $\Psi=0$ on the surface.  While direct calculation of
$\Phi$ in the whole interior
via a volume integration with a Green's function that enforces vacuum BCs
would be computationally very expensive (and would also require
storing large arrays), calculation of $\Theta$ on the surface requires only
integration over the surface.  Once the surface potential $\Phi^{\rm B}$ is known,
it can be used to compute the interior $\Phi$ via an efficient numerical
method.

The DGF and its convolution presented in \citet{james77} are valid for Cartesian coordinates, but it is straightforward to extend James's method to cylindrical coordinates. \citet{snytnikov11} was the first to adapt James's method for 3D cylindrical problems, but he used the CGF in place of DGF. Since the difference between CGF and DGF is quite large at small $\bf |x - x'|$
(see Appendix \ref{s:calc_dgf}), this forced him to make a computational
domain larger than the original volume
in order to improve accuracy.
We note, however, that the domain cannot be extended arbitrarily
across the inner radial boundary $R_{\rm min}$ in cylindrical coordinates, especially when $R_{\rm min}$ is small.  The James algorithm with CGF loses accuracy when the ratio of the outer to inner radial boundary is large, which is frequently encountered in various astronomical applications.
In such situations, it is desirable to use the DGF rather than the CGF for accurate potential computations.

In this paper, we develop an efficient, accurate, and scalable
algorithm to calculate the gravitational potential of an isolated
system in cylindrical coordinates. Our algorithm is fully 3D, and
considers both uniform and logarithmic cylindrical grids.  For
completeness, and to connect with James's original method, we also
present the method for Cartesian coordinates.  Our method utilizes the
James algorithm with the DGF for the boundary potential from screening
charges. For the interior solver,
for cylindrical grids
we employ a hybrid method that incorporates the
eigenfunction expansion in azimuthal and vertical
directions and a tridiagonal solver in the radial direction. For
Cartesian grids our interior solver employs standard Fourier methods.
We  implement our algorithm in the {\tt Athena++} code framework (e.g.,
\citealt{white16}), and parallelize it on a distributed memory
platform using the message passing interface (MPI).  Using various
test problems, we confirm that our method is efficient and
second-order accurate.

The remainder of the paper is organized as follows. In Section \ref{s:equation}, we introduce our 3D computational domain in Cartesian and cylindrical coordinates and discretize the Poisson equation. In Section \ref{s:interior_solver}, we describe the hybrid method
we adopt to solve the discrete Poisson equation in the interior of the computational domain for a given Dirichlet boundary condition.
In Section \ref{s:bc}, we introduce the James algorithm in Cartesian
coordinates and its extension to cylindrical coordinates. In Section \ref{s:tests}, we present the results of our Poisson solver on various test problems to demonstrate its accuracy and efficiency. In Section \ref{s:discussion}, we summarize and discuss the present work.

\section{DISCRETE POISSON EQUATION}\label{s:equation}

A standard way to solve Equation \eqref{eq:Poisson} is a
finite-difference method in which
the differential operator $\nabla^2$ is replaced by the \emph{difference} operator $\Delta^2$, yielding the \emph{discrete} Poisson equation $\Delta^2\Phi = 4\pi G \rho$. The definition of $\Delta^2$ depends on the coordinates and the desired level of approximation. This section defines our computational domains and the finite-difference representations of $\Delta^2$, which are second-order accurate, in uniform Cartesian, uniform cylindrical, and logarithmic cylindrical coordinate systems.

\subsection{Uniform Cartesian Grid}\label{s:uniform_cartesian_grid}
In Cartesian coordinates, we discretize the computational domain $[x_{\rm min},x_{\rm max}]\times[y_{\rm min},y_{\rm max}]\times[z_{\rm min},z_{\rm max}]$ with size $L_x\times L_y\times L_z$ uniformly into $N_x\times N_y\times N_z$ cells. We define the face-centered coordinates as $x_{i+1/2} = x_{\rm min} + i \delta x$ where $\delta x \equiv L_x/N_x$ and similarly for
 $y_{j+1/2}$ and  $z_{k+1/2}$. We also define the cell-centered coordinates as $x_i = (x_{i-1/2} + x_{i+1/2})/2$ with index $i$ running from $1$ to $N_x$, and similarly for $y_j$ with $j=1,2,\cdots, N_y$ and $z_k$ with $k=1,2,\cdots,N_z$.

We denote the cells inside the nominal index range given above as ``active cells", because these are the places where other fluid variables are updated by a hydrodynamics solver. To the active cells, we add one extra layer of ``ghost cells" to the boundaries of the computational domain, with their
cell-centered coordinates are denoted, for example, by $i=0$ and $i=N_x+1$ in the $x$-direction. The boundary conditions for the Poisson equation and other equations of hydrodynamics are provided using these ghost cells. We similarly define the ghost cells in the other coordinate systems described below.

The second-order accurate, finite-difference approximation to Equation \eqref{eq:Poisson} can
be written as
\begin{equation}\label{eq:fd_uniform_cartesian}
  \left(\Delta_x^2 + \Delta_y^2 + \Delta_z^2\right)\Phi_{i,j,k} = 4\pi G \rho_{i,j,k},
\end{equation}
where $\Phi_{i,j,k}$ and $\rho_{i,j,k}$ are the cell-centered potential-density pair and
the difference operators $\Delta_x^2$, $\Delta_y^2$, and $\Delta_z^2$ are defined by
\begin{align}
  \Delta_x^2\Phi_{i,j,k} &= \frac{\Phi_{i-1,j,k}-2\Phi_{i,j,k}+\Phi_{i+1,j,k}}{(\delta x)^2}, \\
  \Delta_y^2\Phi_{i,j,k} &= \frac{\Phi_{i,j-1,k}-2\Phi_{i,j,k}+\Phi_{i,j+1,k}}{(\delta y)^2}, \\
  \Delta_z^2\Phi_{i,j,k} &= \frac{\Phi_{i,j,k-1}-2\Phi_{i,j,k}+\Phi_{i,j,k+1}}{(\delta z)^2}.
  \label{eq:delz2}
\end{align}

\subsection{Uniform Cylindrical Grid}

In uniform cylindrical coordinates, we discretize the computational domain $[R_{\rm min},R_{\rm max}]\times[\phi_{\rm min},\phi_{\rm max}]\times[z_{\rm min},z_{\rm max}]$ with size $L_R\times L_\phi\times L_z$ uniformly into $N_R\times N_\phi\times N_z$ cells. We require that $L_\phi=\phi_{\rm max}-\phi_{\rm min}$ should be an integer fraction of 2$\pi$ to impose periodic boundary condition along the azimuthal direction.
We define the face-centered radial and azimuthal coordinates as $R_{i+1/2} = R_{\rm min} + i \delta R$ and $\phi_{j+1/2} = \phi_{\rm min} + j \delta \phi$, where $\delta R \equiv L_R/N_R$ and $\delta \phi \equiv L_\phi/N_\phi$.
Unlike in Cartesian coordinates, the definition of the radial cell-center is ambiguous in cylindrical coordinates because the geometric center does not coincide with the volumetric center. When the radial grid is uniform, finite difference of quantities defined at the geometric centers can retain second-order accuracy. We thus define the cell-centered coordinates as $R_{i} = (R_{i-1/2} + R_{i+1/2})/2$ with $i=1,2,\cdots,N_R$ and $\phi_{j} = (\phi_{j-1/2} + \phi_{j+1/2})/2$ with $j=1,2,\cdots,N_\phi$. Discretization in the vertical direction is the same as in the uniform Cartesian coordinates.

The second-order finite-difference approximation to Equation \eqref{eq:Poisson} can be written as
\begin{equation}\label{eq:fd_uniform_cylindrical}
  \left( \Delta_R^2 + \Delta_\phi^2 + \Delta_z^2 \right)\Phi_{i,j,k} = 4\pi G \rho_{i,j,k},
\end{equation}
where the difference operators $\Delta_R^2$ and $\Delta_\phi^2$ are defined by
\begin{align}
  \Delta_R^2\Phi_{i,j,k} =& \frac{\Phi_{i-1,j,k}-2\Phi_{i,j,k}+\Phi_{i+1,j,k}}{(\delta R)^2}
    + \frac{\Phi_{i+1,j,k}-\Phi_{i-1,j,k}}{2R_i\delta R},\label{eq:radial_operator_uniform} \\
  \Delta_\phi^2\Phi_{i,j,k} =& \frac{\Phi_{i,j-1,k}-2\Phi_{i,j,k}+\Phi_{i,j+1,k}}{R_i^2(\delta\phi)^2},
\end{align}
while $\Delta_z^2$ is defined through Equation \eqref{eq:delz2}.

\subsection{Logarithmic Cylindrical Grid}

In logarithmic cylindrical coordinates, we discretize a cylindrical computational domain in the same way as in the uniform cylindrical grid, but with logarithmic radial spacing. We define the face-centered radial coordinates as $R_{i+1/2} = f^i R_{\rm min}$, with a common multiplication factor $f\equiv(R_{\rm max} / R_{\rm min})^{1/N_R} >1$. Since the radial zone width, given by $R_{i+1/2}-R_{i-1/2} = (f-1)f^{i-1}R_{\rm min}$ shrinks toward small radii,
a logarithmic cylindrical grid is advantageous in resolving the central regions of a disk with high accuracy. We also define the cell-centered radial coordinates using the volumetric centers as
\begin{equation*}
  R_i \equiv \frac{\int_{R_{i-1/2}}^{R_{i+1/2}} R^2 dR}
             {\int_{R_{i-1/2}}^{R_{i+1/2}} R dR} =
            \frac{2(f^2+f+1)}{3(f+1)} f^{i-1}R_{\rm min},
\end{equation*}
for $i=1,2,\cdots,N_R$. Note that the radial cell spacing $\delta R_i \equiv R_{i+1} - R_i = (f-1)R_i$ increases with $R_i$.

The second-order finite-difference approximation to Equation \eqref{eq:Poisson} takes the same form as Equation \eqref{eq:fd_uniform_cylindrical}, but with the radial difference operator defined as
\begin{equation}\label{eq:radial_operator}
  \Delta_R^2\Phi_{i,j,k} = \frac{\Phi_{i-1,j,k}-2\Phi_{i,j,k}+\Phi_{i+1,j,k}}{(R_i\ln f)^2}.
\end{equation}
Appendix \ref{s:fd} shows that Equation \eqref{eq:radial_operator} makes
the finite difference approximation second-order accurate.

\section{Calculation of Interior Potential for Dirichlet Boundary Conditions}\label{s:interior_solver}

In this section, we provide the general method that we use to obtain the
interior potential within the original domain, given Dirichlet boundaries
for the potential on the surface.  We describe our methods for both Cartesian
grids (Section \ref{sec:cart_int}) and cylindrical grids
(Section \ref{s:interior_solver_cylindrical}).
We note that in principle, alternative fast and efficient
solvers (such as multigrid) could be employed for computing the interior
potential given Dirichlet boundary conditions for the potential.

The interior solver is employed in three different instances in our method.
The first instance is our use of the interior solver to obtain the
numerical DGF, as described in Appendix \ref{s:calc_dgf}; this is done once at the beginning
of any simulation.  The other two instances are in the first and fourth
step  of the James's method; each time the
Poisson solution is required, two calls to the interior solver are made.

In the first step of James's method, the interior potential from the original
density distribution is computed subject to the zero boundary condition.
In the fourth step of James's method,
the interior potential must be computed subject to a
known boundary potential $\Phi^{\rm B}$;
a method to obtain $\Phi^{\rm B}$ will be presented in Section \ref{s:bc}.
Formally, we define $\Phi^{\rm B}$ as having nonzero value only in a single
layer of ghost zones immediately outside the active domain.  Then we
can write the desired potential as $\Phi = \widetilde{\Phi} + \Phi^{\rm B}$ where the
required boundary condition is $\widetilde{\Phi}=0$.  From the definition of
$\Phi^{\rm B}$, $\Delta^2
\Phi^{\rm B}$ will be nonzero only in the
single layer of active zones adjoining the domain boundaries. We can thus define
a modified density distribution $\rho \rightarrow \rho - \Delta^2
\Phi^{\rm B}/(4\pi G)$ which is the same as the original density
distribution everywhere
except in the layer just inside the domain boundaries.  We then
employ this modified density distribution following the procedure of
Section \ref{sec:cart_int} or
Section \ref{s:interior_solver_cylindrical} to compute $\widetilde{\Phi}$.  Within the
interior, where $\Phi^{\rm B}=0$, this solution is then the
desired final solution $\Phi$.

Note that in Sections \ref{sec:cart_int} and
\ref{s:interior_solver_cylindrical} below, $\rho_{i,j,k}$ is any
arbitrary density distribution on the grid, and in fact represents a
different quantity for each of the three instances where we solve for
the interior potential.



\subsection{Cartesian Grid Solution with Zero Boundary Value}\label{sec:cart_int}

It is conventional to utilize the eigenfunctions of a differential operator in solving an elliptic partial differential equation.
The same technique can be applied to the discretized Poisson equation, if the eigenfunctions of the corresponding discrete Laplace operator can be found.

Let ${\cal X}^l_i$, ${\cal Y}^m_j$, and ${\cal Z}^n_k$ be the eigenfunctions of the discrete Laplace operators $\Delta_x^2$, $\Delta_y^2$, and $\Delta_z^2$ satisfying $\Delta_x^2{\cal X}^l_i = \lambda_x^l{\cal X}^l_i$, $\Delta_y^2{\cal Y}^m_j = \lambda_y^m{\cal Y}^m_j$, and $\Delta_z^2{\cal Z}^n_k = \lambda_z^n{\cal Z}^n_k$, with respective eigenvalues $\lambda^l_x$, $\lambda_y^m$, and $\lambda_z^n$.
It is straightforward to show that
\begin{equation}
  {\cal X}^l_i = \sin\left(\frac{\pi li}{N_x+1}\right),\label{eq:car_eigen_x}
\end{equation}
\begin{equation}
  {\cal Y}^m_j = \sin\left(\frac{\pi mj}{N_y+1}\right),\label{eq:car_eigen_y}
\end{equation}
\begin{equation}
  {\cal Z}^n_k = \sin\left(\frac{\pi nk}{N_z+1}\right),\label{eq:car_eigen_z}
\end{equation}
are the desired eigenfunctions satisfying the zero boundary condition at the ghost cells. The corresponding eigenvalues are
\begin{equation}
  \lambda_x^l = -k_l^2\left[ \sin\left( \frac{\pi l}{2(N_x+1)}  \right) \bigg/\left( \frac{\pi l}{2N_x}  \right)  \right]^2,
\end{equation}
\begin{equation}
  \lambda_y^m = -k_m^2\left[ \sin\left( \frac{\pi m}{2(N_y+1)}  \right) \bigg/\left( \frac{\pi m}{2N_y}  \right)  \right]^2,
\end{equation}
\begin{equation}
  \lambda_z^n = -k_n^2\left[ \sin\left( \frac{\pi n}{2(N_z+1)}  \right) \bigg/\left( \frac{\pi n}{2N_z}  \right)  \right]^2,
\end{equation}
where $k_l \equiv \pi l / L_x$, $k_m \equiv \pi m / L_y$, and $k_n \equiv \pi n / L_z$.
In the limit of $l/N_x, m/N_y, n/N_z \ll 1$, the discrete eigenvalues reduce to
the counterpart of the continuous Laplace operator ($-k^2$).

The discrete analog of the Sturm-Liouville theory (e.g., \citealt[chap. 1.10--1.16]{hil68}; see also, \citealt{atk64}) guarantees that the eigenfunctions given in Equation \eqref{eq:car_eigen_x}--\eqref{eq:car_eigen_z} satisfy discrete orthogonality relations, for example,
\begin{equation}\label{eq:orthogorel}
  \frac{2}{N_x+1} \sum_{i=1}^{N_x}{\cal X}_i^l{\cal X}_i^{l'} = \delta_{ll'}\quad\text{and}\quad\frac{2}{N_x+1} \sum_{l=1}^{N_x} {\cal X}_i^l{\cal X}_{i'}^l = \delta_{ii'},
\end{equation}
where the symbol $\delta$ denotes the Kronecker delta. These orthogonality relations allow us to expand
$\widetilde{\Phi}_{i,j,k}$ and $\rho_{i,j,k}$ as
\begin{align}
  \widetilde{\Phi}_{i,j,k} = \frac{8}{(N_x+1)(N_y+1)(N_z+1)}\sum_{l=1}^{N_x}\sum_{m=1}^{N_y}\sum_{n=1}^{N_z}\widetilde{\Phi}^{lmn}{\cal X}_i^l{\cal Y}_j^m{\cal Z}_k^n,\label{eq:car_Phi_forward}\\
  \rho_{i,j,k} = \frac{8}{(N_x+1)(N_y+1)(N_z+1)}\sum_{l=1}^{N_x}\sum_{m=1}^{N_y}\sum_{n=1}^{N_z}\rho^{lmn}{\cal X}_i^l{\cal Y}_j^m{\cal Z}_k^n,\label{eq:car_rho_forward}
\end{align}
where the expansion coefficients $\widetilde{\Phi}^{lmn}$ and $\rho^{lmn}$ are given by
\begin{align}
  \widetilde{\Phi}^{lmn} &= \sum_{i=1}^{N_x}\sum_{j=1}^{N_y}\sum_{k=1}^{N_z}\widetilde{\Phi}_{i,j,k}{\cal X}_i^l{\cal Y}_j^m{\cal Z}_k^n,\label{eq:car_Phi_backward}\\
  \rho^{lmn} &= \sum_{i=1}^{N_x}\sum_{j=1}^{N_y}\sum_{k=1}^{N_z}\rho_{i,j,k}{\cal X}_i^l{\cal Y}_j^m{\cal Z}_k^n.\label{eq:car_rho_backward}
\end{align}
Plugging Equations \eqref{eq:car_Phi_forward}--\eqref{eq:car_rho_forward} in Equation \eqref{eq:fd_uniform_cartesian}, we obtain a simple algebraic relation
\begin{equation}\label{eq:car_Poisson_transform}
  \widetilde{\Phi}^{lmn} = \frac{4\pi G \rho^{lmn}}{\lambda_x^l + \lambda_y^m + \lambda_z^n}.
\end{equation}
Therefore, the Poisson equation in Cartesian coordinates can be solved by the following three steps:
\begin{enumerate}
  \item Perform a forward transform $\rho_{i,j,k}\to\rho^{lmn}$ using Equation \eqref{eq:car_rho_backward} : ${\cal O}(N_xN_yN_z\log_2[N_xN_yN_z])$.
  \item Convert $\rho^{lmn} \to \widetilde{\Phi}^{lmn}$ using the kernel in
    Equation \eqref{eq:car_Poisson_transform} : ${\cal O}(N_xN_yN_z)$.
  \item Perform a backward transform $\widetilde{\Phi}^{lmn}\to\widetilde{\Phi}_{i,j,k}$ using Equation \eqref{eq:car_Phi_forward} : ${\cal O}(N_xN_yN_z\log_2[N_xN_yN_z])$.
\end{enumerate}

In practice, the discrete transforms in Equations \eqref{eq:car_Phi_forward}--\eqref{eq:car_rho_backward} can be performed efficiently with an FFT algorithm. The public {\tt FFTW} library\footnote{\url{http://www.fftw.org/}} performs sine transforms of various kinds, among which we use {\tt FFTW\_R0DFT00} consistent with the zero boundary condition. To perform 3D FFTs in parallel, we decompose the computational domain into 2D pencils along, for example, the $y$-axis and execute 1D sine transforms locally in each pencils (e.g., \citealt{li10}). We then transpose the pencils to the $z$- and $x$-axes sequentially, each time by executing corresponding sine transforms, which completes a 3D FFT. The parallel transpose among different pencil decompositions are done with the {\tt remap\_3d} function in Steve Plimpton's parallel FFT package\footnote{\url{https://www.sandia.gov/~sjplimp/docs/fft/README.html}}. We note that, since mass density and the gravitational potential in general are distributed as blocks rather than pencils in real applications, we have to transpose between block and pencil decompositions at the input and output stage of the Poisson solver. Plimpton's remap routine provides this functionality as well.

\subsection{Cylindrical Grid Solution with Zero Boundary Value}\label{s:interior_solver_cylindrical}

A natural boundary condition in the azimuthal direction is that both mass density and gravitational potential are periodic, with period $L_\phi$. This holds true even when the problem under study has $P$-fold symmetry in the $\phi$-direction, with a domain size $L_\phi = 2\pi / P$. The algorithm presented below is applicable for such systems as long as the $P$-fold  symmetry is considered in the boundary condition for the DGF (Equation \eqref{eq:cyl_asymptotic_green}).

The eigenfunction ${\cal P}^m_j$ for the discrete Laplace operator $\Delta_\phi^2$ and the corresponding eigenvalue $\lambda_\phi^m$ are given by
\begin{equation}\label{eq:cyl_eigenfunction}
  {\cal P}^m_j = \exp\left[ \frac{2\pi\sqrt{-1}mj}{N_\phi}  \right],
\end{equation}
\begin{equation}
  \lambda^m_\phi = -\frac{m^2}{R_i^2}\left[ \sin\left( \frac{\pi m}{N_\phi}  \right) \bigg/\left( \frac{\pi m}{N_\phi}  \right)  \right]^2.
\end{equation}
Note that $\lambda^m_\phi \rightarrow -m^2/R^2$ for $m/N_\phi \ll 1$.
The eigenfunction ${\cal P}^m_j$ satisfies the discrete orthogonality relation
\begin{equation}
  \frac{1}{N_\phi} \sum_{j=1}^{N_\phi}({\cal P}_j^m)^* {\cal P}^{m'}_j = \delta_{mm'}\quad\text{and}\quad\frac{1}{N_\phi} \sum_{m=1}^{N_\phi}({\cal P}_j^m)^* {\cal P}^{m'}_j = \delta_{jj'}.
\end{equation}

We expand $\widetilde{\Phi}_{i,j,k}$ and $\rho_{i,j,k}$ only along the
azimuthal and vertical directions as
\begin{align}
  \widetilde{\Phi}_{i,j,k} = \frac{2}{N_\phi (N_z+1)}\sum_{m=1}^{N_\phi}\sum_{n=1}^{N_z}\widetilde{\Phi}^{mn}_{i}{\cal P}_j^m{\cal Z}_k^n,\label{eq:cyl_Phi_forward}\\
  \rho_{i,j,k} = \frac{2}{N_\phi (N_z+1)}\sum_{m=1}^{N_\phi}\sum_{n=1}^{N_z}\rho^{mn}_{i}{\cal P}_j^m{\cal Z}_k^n,\label{eq:cyl_rho_forward}
\end{align}
where the expansion coefficients $\widetilde{\Phi}^{mn}_{i}$ and $\rho^{mn}_{i}$ satisfy the inverse transforms
\begin{align}
  \widetilde{\Phi}^{mn}_{i} &= \sum_{j=1}^{N_\phi}\sum_{k=1}^{N_z}\widetilde{\Phi}_{i,j,k} ({\cal P}^m_j)^* {\cal Z}^n_k,\label{eq:cyl_Phi_backward}\\
  \rho^{mn}_{i} &= \sum_{j=1}^{N_\phi}\sum_{k=1}^{N_z}\rho_{i,j,k} ({\cal P}^m_j)^* {\cal Z}^n_k.\label{eq:cyl_rho_backward}
\end{align}
One cannot analytically expand $\widetilde{\Phi}_{i,j,k}$ and $\rho_{i,j,k}$ along the radial direction because radial eigenfunction ${\cal R}^l_i$, defined through $\Delta_R^2{\cal R}^l_i = \lambda^l_R {\cal R}^l_i$, has no closed-form expression and is not compatible with FFT.\footnote{The radial eigenfunction ${\cal R}^l_i$ can instead be obtained numerically by solving the eigenvalue problem, and the resulting eigenfunction may be called the \emph{discrete Bessel function}. Since it satisfies the exact discrete orthogonality relation, it may also serve as discrete kernel for the discrete Hankel transform \citep{john87,baddour15}.}

Plugging Equation \eqref{eq:cyl_Phi_forward}--\eqref{eq:cyl_rho_forward} into Equation \eqref{eq:fd_uniform_cylindrical} yields
\begin{equation}\label{eq:tridiagonal}
  \left( \Delta_R^2 + \lambda^m_\phi + \lambda^n_z \right) \widetilde{\Phi}^{mn}_i = 4\pi G \rho^{mn}_i,
\end{equation}
which, using Equations \eqref{eq:radial_operator_uniform} and \eqref{eq:radial_operator}, can be written as
\begin{equation}\label{eq:tridiagonal_uni}
  \left[ \frac{1}{(\delta R)^2} - \frac{1}{2R_i \delta R}  \right] \widetilde{\Phi}_{i-1}^{mn} + \left[ \lambda_\phi^m + \lambda^n_z - \frac{2}{(\delta R)^2}  \right] \widetilde{\Phi}_i^{mn} + \left[ \frac{1}{(\delta R)^2} + \frac{1}{2R_i \delta R}  \right] \widetilde{\Phi}_{i+1}^{mn} = 4\pi G \rho^{mn}_i
\end{equation}
in uniform cylindrical coordinates, and
\begin{equation}\label{eq:tridiagonal_log}
  \frac{1}{(R_i\ln f)^2}  \widetilde{\Phi}_{i-1}^{mn} + \left[ \lambda_\phi^m + \lambda^n_z - \frac{2}{(R_i\ln f)^2}  \right] \widetilde{\Phi}_i^{mn} + \frac{1}{(R_i\ln f)^2} \widetilde{\Phi}_{i+1}^{mn} = 4\pi G \rho^{mn}_i
\end{equation}
in logarithmic cylindrical coordinates. Note that
Equations \eqref{eq:tridiagonal_uni} and \eqref{eq:tridiagonal_log}
are tridiagonal matrix equations subject to the
zero boundary conditions of $\widetilde{\Phi}_{0}^{mn} = \widetilde{\Phi}_{N_R+1}^{mn} = 0$,
which can easily be solved via the Thomas algorithm involving back substitutions
(e.g., \citealt{nr}).

Therefore, the Poisson equation in cylindrical coordinates can be solved by the following
three steps:
\begin{enumerate}
  \item Perform a forward transform $\rho_{i,j,k}\to\rho^{mn}_i$ using Equation \eqref{eq:cyl_rho_backward} : ${\cal O}(N_RN_\phi N_z\log_2[N_\phi N_z])$.
  \item Solve Equation \eqref{eq:tridiagonal_uni} or
    \eqref{eq:tridiagonal_log} for $\widetilde{\Phi}^{mn}_i$  : ${\cal O}(N_RN_\phi N_z)$.
  \item Perform a backward transform $\widetilde{\Phi}^{mn}_i\to\widetilde{\Phi}_{i,j,k}$ using Equation \eqref{eq:cyl_Phi_forward} : ${\cal O}(N_RN_\phi N_z\log_2[N_\phi N_z])$.
\end{enumerate}

In actual computation, the discrete transforms in Equations \eqref{eq:cyl_Phi_forward}--\eqref{eq:cyl_rho_backward} can be carried out efficiently with an FFT algorithm. For transforms involving ${\cal P}^m_j$, we use the real-to-complex transform in {\tt FFTW}, which halves the size of the output by utilizing the Hermitian symmetry. For transforms involving ${\cal Z}^n_k$, we use the sine transform as in Cartesian coordinates. For parallel computations, we employ the pencil decomposition technique along with the Steve Plimpton's parallel transpose routines, similarly to the Cartesian solver.

\section{CALCULATION OF THE BOUNDARY POTENTIAL}\label{s:bc}

\subsection{Overview of the James Algorithm}\label{s:James_overview}

We adopt the James algorithm to calculate the boundary potential $\Phi^{\rm B}$
which is second-order accurate. As explained in Introduction, James's method first solves for the preliminary potential $\Psi=\Phi+\Theta$ with zero boundary condition, where $\Phi({\bf x}) = -\int G\rho({\bf x'})/|{\bf x-x'}|\,d^3x'$ is the gravitational potential generated from the original density distribution (i.e., the desired solution) and $\Theta({\bf x}) = -\oint G\sigma({\bf x'})/|{\bf x-x'}|\,d^2x'$ is that from the screening charges. Since $\Psi^{\rm B}=0$ by definition, one only needs to compute $\Theta^{\rm B}=-\oint G\sigma({\bf x'})/|{\bf x}^{\rm B} - {\bf x'}|\,d^2x'$ to obtain $\Phi^{\rm B} = -\Theta^{\rm B}$.

\begin{figure}[b!]
  \epsscale{0.6}
  \plotone{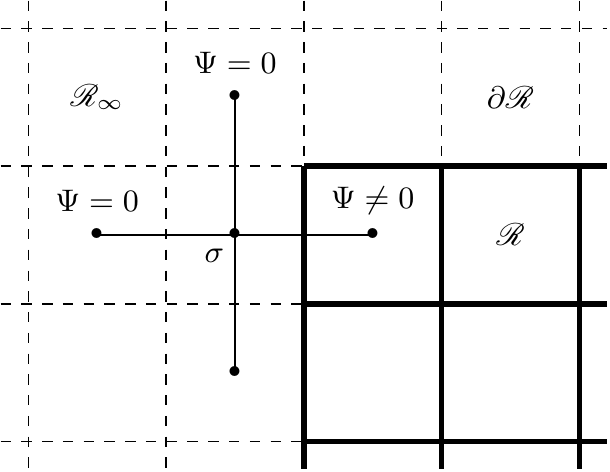}
  \caption{Schematic diagram describing the screening charge calculation. $\mathscr R$ is the original computational domain and $\partial{\mathscr R}$ denotes the domain boundary or the ghost zones. $\mathscr{R}_\infty$ represents the hypothetical infinite discrete domain, where $\Psi=0$ but $\Phi$ and $\Theta$ are nonzero.}
  \label{fig:james}
\end{figure}

Once the preliminary gravitational potential $\Psi$ with the zero boundary condition is obtained (using the method of Section \ref{s:interior_solver} and the original density distribution), the screening charges are found by applying the discrete Laplace operators at the ghost cells. It should be noted that at the ghost cells, the discrete Laplace operator calls for the value of $\Psi$ \emph{outside} the ghost cells, which is set to zero in \citet{james77}. Figure \ref{fig:james} depicts this situation, where ${\mathscr R}$ and $\partial{\mathscr R}$ represent the active and the ghost cells defined in Section \ref{s:equation}. Encompassing this, one can imagine the infinite domain ${\mathscr R}_\infty$ where $\Psi=\rho=0$ everywhere exterior to $\partial{\mathscr R}$. It is evident that the $\Psi$ satisfies the discrete Poisson equation at every cell in ${\mathscr R}_\infty$ if one adds $\sigma=\Delta^2\Psi/(4\pi G)$ at $\partial{\mathscr R}$. Since $\Delta^2\Psi = 4\pi G (\rho+\sigma)$ in $\mathscr R_\infty$, one can solve $\Delta^2\Theta = 4\pi G\sigma$ to obtain $\Phi = \Psi - \Theta$ that satisfies the original Poisson equation $\Delta^2\Phi = 4\pi G\rho$ subject to the vacuum boundary condition (i.e., $\Phi$ is the desired solution for an isolated mass distribution). Note that one needs $\Theta$ only at the domain boundary, which can be efficiently calculated using the Green's function for the discrete Laplace operator subject to vacuum boundary condition (i.e., the DGF). Then, $\Phi^{\rm B} = -\Theta^{\rm B}$ gives a new boundary condition for the final step of the James's method.

To summarize, the James algorithm consists of the following four steps: (1) Solve the Poisson equation with
the zero boundary condition to obtain $\Psi$; (2) Evaluate the screening charge $\sigma$ by applying the discrete Laplace operator to the ghost cells ($\sigma = \Delta^2\Psi/4\pi G$); (3) Use the DGF to calculate
the gravitational potential $\Theta^{\rm B}$ at the domain boundary due to $\sigma$;
(4) Solve the Poisson equation with the Dirichlet boundary condition
$\Phi^{\rm B} = - \Theta^{\rm B}$.
In section \ref{s:interior_solver} we have presented the method we adopt
for the interior Poisson solver, which is employed in Steps (1) and (4) and
also is used to pre-compute the DGF (see Appendix \ref{s:calc_dgf}), which enters in Step 3.
In what follows, we describe Steps (2) and (3) in more detail.

\subsection{Computation of the Screening Charges}\label{s:James_boundary_charges}

Once the preliminary gravitational potential $\Psi$ with the zero boundary condition is obtained (using the method of Section \ref{s:interior_solver} and
the original density distribution),
one can readily apply the discrete Laplace operators at the ghost cells in each boundary to calculate the screening charges.

\subsubsection{Cartesian Grid}

A Cartesian grid has six boundary surfaces consisting of the loci of ghost
zones immediately outside the problem domain:
bottom (bot; $k=0$), top (top; $k=N_z+1$),
south (sth; $j=0$), north (nth; $j=N_y+1$), west (wst; $i=0$), and east (est; $i=N_x+1$).
With $\Psi=0$ in both the first and second layer of ghost zones outside
the domain, the screening charges on these boundary surfaces are given by
\begin{align}
  \sigma_{i,j}({\rm bot}) &= \frac{1}{4\pi G (\delta z)^2} \Psi|_{k=1}, & \sigma_{i,j}({\rm top}) &= \frac{1}{4\pi G (\delta z)^2} \Psi|_{k=N_z},\nonumber\\
  \sigma_{j,k}({\rm wst}) &= \frac{1}{4\pi G (\delta x)^2} \Psi|_{i=1}, & \sigma_{j,k}({\rm est}) &= \frac{1}{4\pi G (\delta x)^2} \Psi|_{i=N_x},\nonumber\\
  \sigma_{i,k}({\rm sth}) &= \frac{1}{4\pi G (\delta y)^2} \Psi|_{j=1}, & \sigma_{i,k}({\rm nth}) &= \frac{1}{4\pi G (\delta y)^2} \Psi|_{j=N_y},
\end{align}
where $\sigma_{i,j}({\rm bot})$ denotes the screening charge on the bottom boundary, etc. Note that the screening charges have units of mass density rather than surface density,
because the charge is assumed to fill a volume $\delta x \delta y \delta z$.

\subsubsection{Cylindrical Grid}

In cylindrical coordinates, one needs to deal with only four boundary surfaces: bottom (bot; $k=0$), top (top; $k=N_z+1$), inner (inn; $i=0$), and outer (out; $i=N_R+1$): the azimuthal direction is assumed periodic.
Using the discrete Laplace operators given in Section \ref{s:equation}, one can show that the screening charges are calculated as
\begin{align}
  \sigma_{i,j}({\rm bot}) &= \frac{1}{4\pi G (\delta z)^2} \Psi|_{k=1}, & \sigma_{i,j}({\rm top}) &= \frac{1}{4\pi G (\delta z)^2} \Psi|_{k=N_z},\nonumber\\
  \sigma_{j,k}({\rm inn}) &= \frac{1+\delta R/(2R_{0})}{4\pi G (\delta R)^2} \Psi|_{i=1}, & \sigma_{j,k}({\rm out}) &= \frac{1-\delta R/(2R_{N_R+1})}{4\pi G (\delta R)^2} \Psi|_{i=N_R}
\end{align}
in a uniform cylindrical grid, and
\begin{align}
  \sigma_{i,j}({\rm bot}) &= \frac{1}{4\pi G (\delta z)^2} \Psi|_{k=1}, & \sigma_{i,j}({\rm top}) &= \frac{1}{4\pi G (\delta z)^2} \Psi|_{k=N_z},\nonumber\\
  \sigma_{j,k}({\rm inn}) &= \frac{1}{4\pi G (R_{0}\ln f)^2} \Psi|_{i=1}, & \sigma_{j,k}({\rm out}) &= \frac{1}{4\pi G (R_{N_R+1}\ln f)^2} \Psi|_{i=N_R}
\end{align}
in a logarithmic cylindrical grid.

\subsection{Discrete Green's Function and the Potential Generated by Screening Charges}\label{s:dgf}

The gravitational potential $\Theta$ that results from the screening charges can be obtained by convolving $\sigma$ with the Green's function of the operator that determines $\sigma$. Since $\sigma$ is obtained through the application of the \emph{discrete} Laplace operator, the corresponding Green's function should be the DGF rather than the CGF. The proper operation of the James's method thus relies on the accurate calculation of the DGF, yet
finding its analytic expressions in cylindrical coordinates is a daunting task.
To our knowledge, the analytic DGF is available only in 2D Cartesian coordinates \citep{bune71}. In 3D Cartesian coordinates, \citet{burk97} addressed the definition, existence, and uniqueness of the DGF and derived asymptotic expansion formulae, applicable at distances far from the source.
In this section, we provide a working definition of the DGF and a numerical method to calculate $\Theta$ in Cartesian and cylindrical coordinates. We refer the reader to Appendix \ref{s:calc_dgf} for our method for the DGF.

\subsubsection{Cartesian Grid}

The DGF, ${\cal G}_{i-i',j-j',k-k'}$, in Cartesian coordinates is the gravitational potential per unit mass due to a discrete point mass at $(i',j',k')$
and ought to satisfy
\begin{equation}\label{eq:def_car_green}
  \left(\Delta_x^2 + \Delta_y^2 + \Delta_z^2\right){\cal G}_{i-i',j-j',k-k'} = 4\pi G \frac{\delta_{ii'}\delta_{jj'}\delta_{kk'}}{\cal V},
\end{equation}
where the symbol $\delta_{ii'}$ is the Kronecker delta and ${\cal V}=\int_{z_{k'-1/2}}^{z_{k'+1/2}}\int_{y_{j'-1/2}}^{y_{j'+1/2}}\int_{x_{i'-1/2}}^{x_{i'+1/2}}dxdydz = \delta x\delta y\delta z$ is the volume of the $(i',j',k')$-th cell. Note that in writing the indices of
${\cal G}_{i-i',j-j',k-k'}$,
we implicitly allow for the translational symmetry on a Cartesian grid.
In Appendix \ref{s:calc_dgf_cart}, we follow \citet{james77} to
calculate the Cartesian DGF numerically.

The gravitational potential $\Theta_{i,j,k}$ generated by the screening charges $\sigma_{i,j,k}$
is given by
\begin{equation}\label{eq:car_bpot_by_dgf}
  \Theta_{i,j,k} = \sum_{i'=0}^{N_x+1}\sum_{j'=0}^{N_y+1}\sum_{k'=0}^{N_z+1}{\cal G}_{i-i',j-j',k-k'}\sigma_{i',j',k'}{\cal V}.
\end{equation}
Substituting Equation \eqref{eq:car_bpot_by_dgf}
into Equation \eqref{eq:fd_uniform_cartesian},
one can easily check that $\Theta_{i,j,k}$ and $\sigma_{i,j,k}$ is a valid potential-density pair.

Because Equation \eqref{eq:car_bpot_by_dgf} involves a discrete convolution,
it is efficient to use FFTs for computations.
Making use of certain symmetries of the problem for a hollow  charge
distribution,
\citet{james77}
devised a formulation that uses sine and cosine transforms
to expresses  the potential on the each surface as the sum of seven
terms.
We refer the reader to Equations (4.7)--(4.20) of \citet{james77}
for the description of this formulation, which costs ${\cal O}(N^3 + N^2 \log N)$ operations.

The gravitational potential $\Phi^{\rm B}$ at the domain boundary is obtained by
\begin{equation}
  \Phi^{\rm B}_{i,j,k} = \Psi^{\rm B}_{i,j,k} - \Theta^{\rm B}_{i,j,k} = -\Theta^{\rm B}_{i,j,k},
\end{equation}
which provides the required Dirichlet boundary condition for the interior solver (Section \ref{s:interior_solver}). Note that $\Psi^{\rm B}_{i,j,k} = 0$ by definition.

\subsubsection{Cylindrical Grid}

The cylindrical DGF, ${\cal G}_{i,i',j-j',k-k'}$, satisfies
\begin{equation}\label{eq:def_cyl_green}
  \left(\Delta_R^2 + \Delta_\phi^2 + \Delta_z^2\right){\cal G}_{i,i',j-j',k-k'} = 4\pi G \frac{\delta_{ii'}\delta_{jj'}\delta_{kk'}}{{\cal V}_{i'}},
\end{equation}
where ${\cal V}_{i'}=\int_{z_{k'-1/2}}^{z_{k'+1/2}}\int_{\phi_{j'-1/2}}^{\phi_{j'+1/2}}\int_{R_{i'-1/2}}^{R_{i'+1/2}} R\,dR d\phi dz = \tfrac{1}{2}(R_{i'+1/2}^2-R_{i'-1/2}^2)\delta\phi\delta z$ is the volume of the $(i',j',k')$-th cell.
Note that the cylindrical DGF has four indices due to lack of the translational symmetry along the radial direction. In Appendix \ref{s:calc_dgf_cyl}, we present the method to calculate the cylindrical DGF and compare it with the continuous counterpart.

The gravitational potential $\Theta_{i,j,k}$ generated by the screening charges $\sigma_{i,j,k}$ takes a form of
\begin{equation}\label{eq:gpot_by_discrete_green}
  \Theta_{i,j,k} = \sum_{i'=0}^{N_R+1}\sum_{j'=1}^{N_\phi}\sum_{k'=0}^{N_z+1} {\cal G}_{i,i',j-j',k-k'}\sigma_{i',j',k'}{\cal V}_{i'}.
\end{equation}
One can readily verify that $\Theta_{i,j,k}$ satisfies the discrete Poisson equation in cylindrical coordinates. Since all functions are periodic in the azimuthal direction, it is natural to apply a discrete Fourier transform such that
\begin{equation}\label{eq:dft}
  \Theta^m_{ik} \equiv \sum_{j=1}^{N_\phi} \Theta_{ijk} e^{-2\pi \sqrt{-1}jm/N_\phi},
\end{equation}
and similarly for ${\cal G}$ and $\sigma$. Then, Equation \eqref{eq:gpot_by_discrete_green} can be cast into a more compact form
\begin{equation}\label{eq:gpot_by_discrete_green_fft}
  \Theta^m_{ik} = \sum_{i'=0}^{N_R+1}\sum_{k'=0}^{N_z+1}{\cal G}^m_{i,i',k-k'} \sigma^m_{i'k'}{\cal V}_{i'}.
\end{equation}
Since $\sigma$ is nonzero only at the ghost cells and we need to evaluate $\Theta$ also only at the ghost cells, we do not have to perform full double summations for all $m,i,k$ indices in Equation \eqref{eq:gpot_by_discrete_green_fft}. By collecting the individual contributions of the surface charges, Equation \eqref{eq:gpot_by_discrete_green_fft} yields the Fourier-transformed potentials
at the four boundaries
\begin{align}\label{eq:Theta_top}
  \Theta^m_i({\rm top}) &= \sum_{i'=1}^{N_R} {\cal G}^m_{i,i'}({\rm top\to top}) \sigma^m_{i'}({\rm top}){\cal V}_{i'} + \sum_{i'=1}^{N_R} {\cal G}^m_{i,i'}({\rm bot\to top}) \sigma^m_{i'}({\rm bot}){\cal V}_{i'}\nonumber\\
                        &+ \sum_{k'=1}^{N_z} {\cal G}^m_{i,k'}({\rm inn\to top}) \sigma^m_{k'}({\rm inn}){\cal V}_{0} + \sum_{k'=1}^{N_z} {\cal G}^m_{i,k'}({\rm out\to top}) \sigma^m_{k'}({\rm out}){\cal V}_{N_R+1},
\end{align}
\begin{align}\label{eq:Theta_bot}
  \Theta^m_i({\rm bot}) &= \sum_{i'=1}^{N_R} {\cal G}^m_{i,i'}({\rm top\to bot}) \sigma^m_{i'}({\rm top}){\cal V}_{i'} + \sum_{i'=1}^{N_R} {\cal G}^m_{i,i'}({\rm bot\to bot}) \sigma^m_{i'}({\rm bot}){\cal V}_{i'}\nonumber\\
                        &+ \sum_{k'=1}^{N_z} {\cal G}^m_{i,k'}({\rm inn\to bot}) \sigma^m_{k'}({\rm inn}){\cal V}_{0} + \sum_{k'=1}^{N_z} {\cal G}^m_{i,k'}({\rm out\to bot}) \sigma^m_{k'}({\rm out}){\cal V}_{N_R+1},
\end{align}
\begin{align}\label{eq:Theta_inn}
  \Theta^m_k({\rm inn}) &= \sum_{i'=1}^{N_R} {\cal G}^m_{k,i'}({\rm top\to inn}) \sigma^m_{i'}({\rm top}){\cal V}_{i'} + \sum_{i'=1}^{N_R} {\cal G}^m_{k,i'}({\rm bot\to inn}) \sigma^m_{i'}({\rm bot}){\cal V}_{i'}\nonumber\\
                        &+ \sum_{k'=1}^{N_z} {\cal G}^m_{k-k'}({\rm inn\to inn}) \sigma^m_{k'}({\rm inn}){\cal V}_{0} + \sum_{k'=1}^{N_z} {\cal G}^m_{k-k'}({\rm out\to inn}) \sigma^m_{k'}({\rm out}){\cal V}_{N_R+1},
\end{align}
\begin{align}\label{eq:Theta_out}
  \Theta^m_k({\rm out}) &= \sum_{i'=1}^{N_R} {\cal G}^m_{k,i'}({\rm top\to out}) \sigma^m_{i'}({\rm top}){\cal V}_{i'} + \sum_{i'=1}^{N_R} {\cal G}^m_{k,i'}({\rm bot\to out}) \sigma^m_{i'}({\rm bot}){\cal V}_{i'}\nonumber\\
                        &+ \sum_{k'=1}^{N_z} {\cal G}^m_{k-k'}({\rm inn\to out}) \sigma^m_{k'}({\rm inn}){\cal V}_{0} + \sum_{k'=1}^{N_z} {\cal G}^m_{k-k'}({\rm out\to out}) \sigma^m_{k'}({\rm out}){\cal V}_{N_R+1},
\end{align}
where we use symbolic notations such that ${\cal G}^m_{i,k'}({\rm inn\to top}) = {\cal G}^m_{i,0,N_z+1-k'}$, $\sigma^m_{i'}({\rm top}) = \sigma^m_{i',N_z+1}$, etc.

Finally, we apply an inverse Fourier transform to $\Theta^m_i({\rm top}), \Theta^m_i({\rm bot}), \Theta^m_k({\rm inn})$, and $\Theta^m_k({\rm out})$ to obtain the boundary potential $\Theta_{i,j,k}^{\rm B}$
due to the surface charges. Then, the desired boundary potential
$\Phi^{\rm B}$ due to the original charge $\rho$ is given by
\begin{equation}\label{eq:boundary_condition}
  \Phi^{\rm B}_{i,j,k} = \Psi^{\rm B}_{i,j,k} -\Theta_{i,j,k}^{\rm B} = -\Theta_{i,j,k}^{\rm B},
\end{equation}
which gives the required Dirichlet boundary condition for the interior solver. Note that the boundary potential calculations explained above involve FFTs on 2D arrays (e.g., $\sigma^m_{i'}({\rm top})$) together with the summations of the Green's function amounting to ${\cal O}(N^3)$ operations. Therefore, the overall computational cost of the boundary potential calculation is of order ${\cal O}(N^3 + N^2\log N)$, similarly to the case with a Cartesian grid.

We note that the above formulation for the boundary potential is valid for a mass distribution under $P$-fold symmetry in $\phi$. In Appendix \ref{s:P-fold_symm}, we directly demonstrate that this is really the case
as long as ${\cal G}_{i,i',j-j',k-k'}$ in Equation
\eqref{eq:gpot_by_discrete_green} properly accounts for the contributions
to the boundary potential from all periodic images of the mass density.

\section{TEST RESULTS}\label{s:tests}

We implement our Poisson solver in {\tt Athena++} which is a state-of-art astrophysical magnetohydrodynamics (MHD) code with
very flexible coordinate and grid options. Using Cartesian and uniform/logarithmic cylindrical grids, we test our solver on a few test problems to check its accuracy, convergence, and parallel performance. We also run time-dependent simulations of a gravitationally-unstable isothermal ring to check if the gravity module combines well with the MHD solver of {\tt Athena++} to produce the expected results of ring fragmentation.
For all tests presented below, we set the gravitational constant to $G = 1$.

\subsection{Uniform Sphere Test}\label{s:staticPot}

To test the accuracy of our Poisson solver, we consider a uniform sphere with radius $r_0$ and density $\rho_0$. The analytic gravitational potential of such a sphere is given by
\begin{equation}
  \Phi_a(r) =
  \begin{dcases}
    -2\pi G \rho_0 \left(r_0^2 - \tfrac{1}{3} r^2\right), & (r < r_0),\\
    -\frac{4\pi G\rho_0 r_0^3}{3r}, & (r > r_0),
  \end{dcases}
\end{equation}
where $r$ denotes the distance from the center of the sphere. We take $r_0=0.2$ and $\rho_0=1$, and place it at an off-centered position
$(x_0, y_0, z_0)$ in Cartesian coordinates and
$(R_0, \phi_0, z_0)$ in cylindrical coordinates, and calculate the gravitational potential $\Phi$ numerically. Table \ref{tb:sphere_test} lists the grid dimension, resolution, and sphere position in  each coordinate system adopted.

\begin{figure*}[t!]
  \plotone{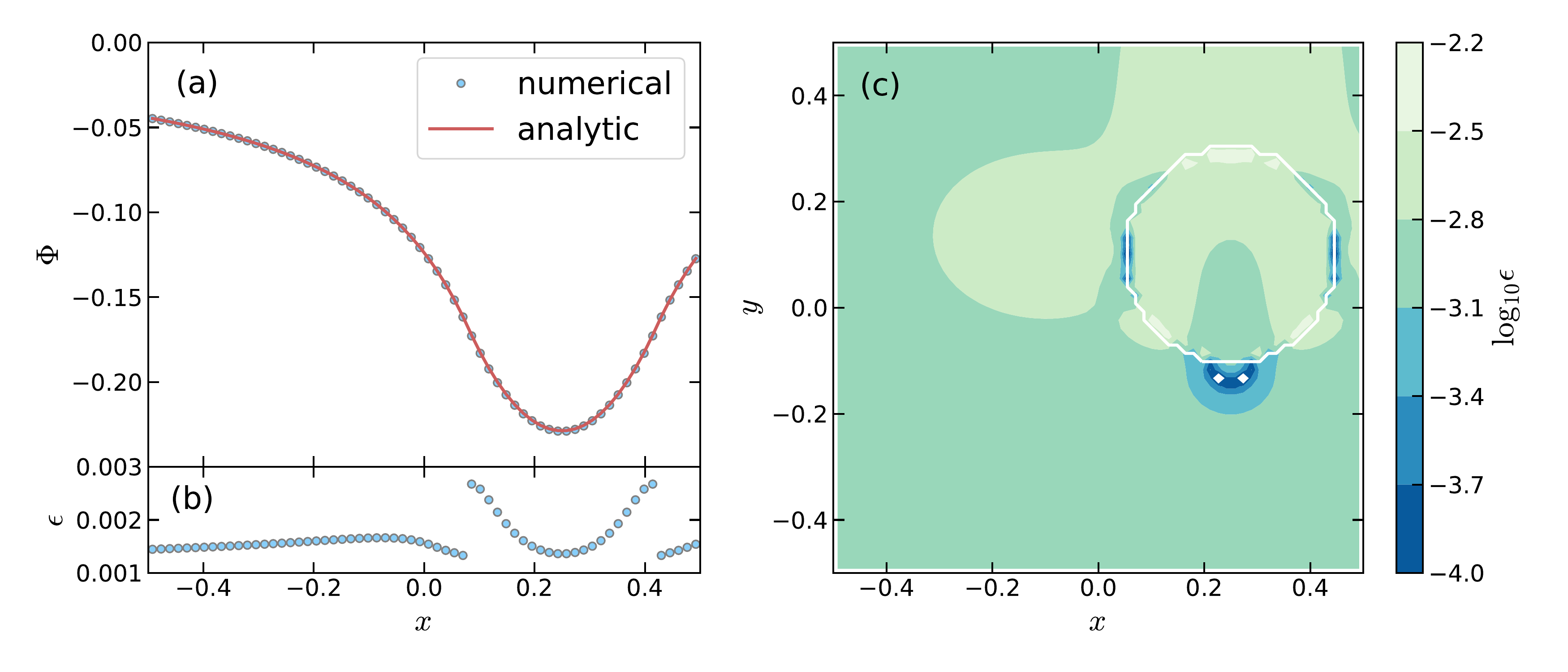}
  \caption{Uniform sphere test on a Cartesian grid. (a) The numerical potential (circles) in comparison with the analytic potential (line) and (b) the relative errors along the $x$-direction at $y=0.00781$ ($j=33$) and $z=0.00781$ ($k=33$). (c) The contour of the relative errors in the midplane ($k=33$), with a white line delineating the sphere boundary on the adopted grid.}
  \label{fig:car_sphere}
\end{figure*}

\begin{figure*}[t!]
  \plotone{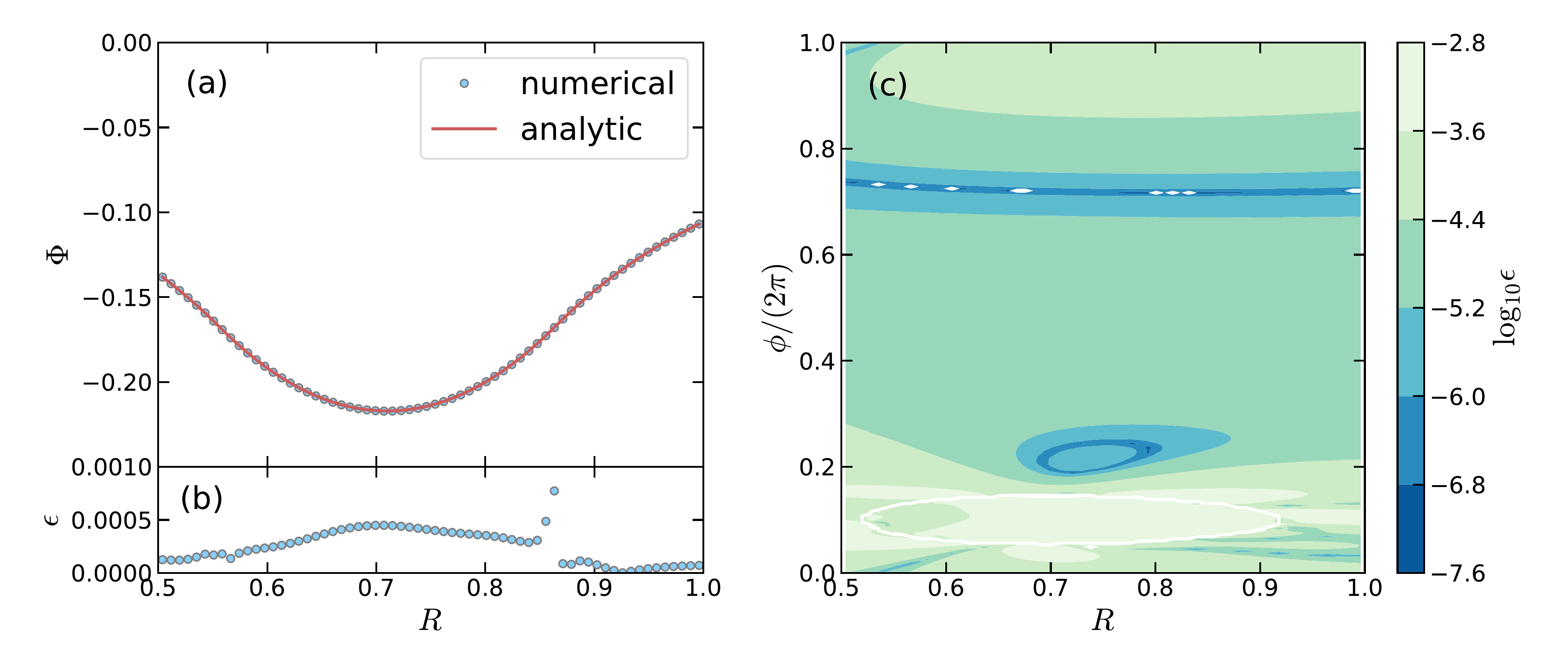}
  \caption{Uniform sphere test on a uniform cylindrical grid. (a) The numerical potential (circles) in comparison with the analytic potential and (b) the relative errors along the $R$-direction at $\phi=0.798$ ($j=33$) and $z=0.00391$ ($k=33$). (c) The contour of the relative error in the midplane ($k=33$), with a white line delineating the sphere boundary on the adopted grid.}
  \label{fig:ucyl_sphere}
\end{figure*}

\begin{figure*}[htb!]
 \plotone{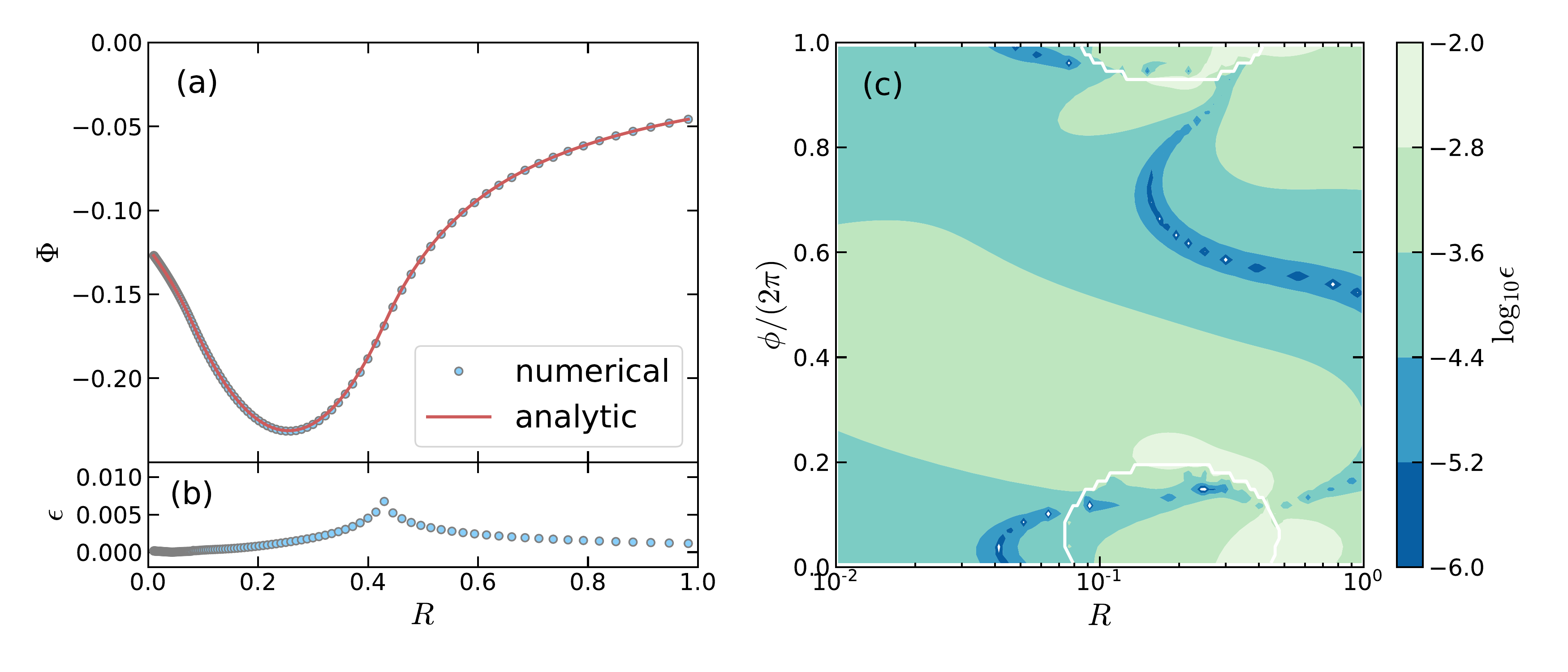}
  \caption{Uniform sphere test on a logarithmic cylindrical grid. (a) The numerical potential (circles) in comparison with the analytic potential and (b) the relative errors along the $R$-direction at $\phi=0.0491$ ($j=1$) and $z=0.00391$ ($k=33$). (c) The contour of the relative error in the midplane ($k=33$), with a white line delineating the sphere boundary on the adopted grid.}
  \label{fig:cyl_sphere}
\end{figure*}

As a measure of accuracy, we define the relative error between the numerical solution and the analytic solution as
\begin{equation}
  {\epsilon} \equiv \left|\frac{\Phi-\Phi_a}{\Phi_a}\right|,
\end{equation}
evaluated at the cell centers.
Figures \ref{fig:car_sphere}--\ref{fig:cyl_sphere} plot the test results on the Cartesian, uniform cylindrical, and logarithmic cylindrical grid, respectively. Panels (a) and (b) plot the one-dimensional (1D)  cut profiles of $\Phi$ and $\epsilon$ along the $x$- or $R$-direction, respectively, while panel (c) gives the 2D distribution of $\epsilon$ in the $z=0$ plane. Overall, the numerical results are exceedingly close to the analytic  potential, with the mean relative error less than 0.1\%. The errors are largest near the sphere boundary whose exact shape is not well resolved by any of the adopted grids.

\begin{deluxetable}{cccccc}[!t]
  \tablecaption{Parameters for the uniform sphere test \label{tb:sphere_test}}
  \tablehead{
    \colhead{Coordinate System} & \colhead{Domain Size} & \colhead{Resolution} & \colhead{$x_0$ or $R_0$} & \colhead{$y_0$ or $\phi_0$} & \colhead{$z_0$}
  }
  \startdata
  Cartesian & $[-0.5,0.5]\times[-0.5,0.5]\times[-0.5,0.5]$ & $64\times 64\times 64$ & $0.25$ & $0.1$ & $-0.04$ \\
  uniform cylindrical & $[0.5,1]\times[0,2\pi]\times[-0.25,0.25]$ & $64\times 256\times 64$ & $0.72$ & $0.63$ & $-0.04$\\
  logarithmic cylindrical & $[10^{-2},1]\times[0,2\pi]\times[-0.25,0.25]$ & $128\times 64\times 64$ & $0.27$ & $0.38$ & $-0.04$
  \enddata
  \tablecomments{The domain size indicates $[x_{\rm min}, x_{\rm max}]\times [y_{\rm min}, y_{\rm max}]\times [z_{\rm min}, z_{\rm max}]$ in the Cartesian grid and $[R_{\rm min},R_{\rm max}]\times [\phi_{\rm min},\phi_{\rm max}]\times [z_{\rm min},z_{\rm max}]$ in the cylindrical grids. Similarly, the resolution indicates $N_x\times N_y\times N_z$ in the Cartesian grid and $N_R\times N_\phi\times N_z$ in the cylindrical grids.}
\end{deluxetable}

\subsection{Convergence Test}\label{s:convergence}

\begin{figure}[htb!]
 \plotone{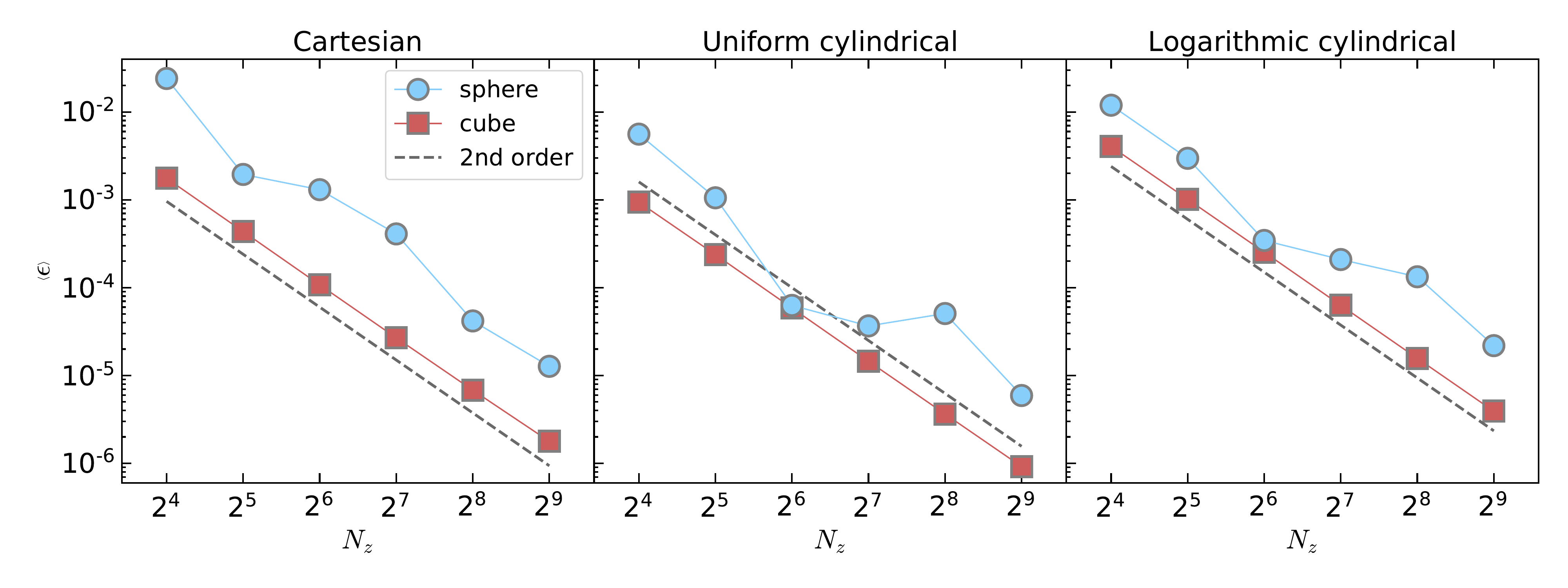}
  \caption{Convergence test results in Cartesian (\emph{left}), uniform cylindrical (\emph{middle}), logarithmic cylindrical grid (\emph{right}). The circles
are the mean relative errors for the uniform sphere, while
the squares are for the uniform cube in the Cartesian grid
and the uniform rectangular torus in the cylindrical grids.
The number of cells is $N_x=N_y=N_z$ for the Cartesian grid, $N_R=N_\phi/4=N_z$ for the uniform cylindrical grid, and $N_R/2=N_\phi=N_z$ for the logarithmic cylindrical grid. In each panel, the dashed line with $-2$ slope is shown for comparison.}
  \label{fig:convergence}
\end{figure}

The discrete Poisson equation used for the interior solver in Section \ref{s:interior_solver} and for the boundary condition in Section \ref{s:bc} are second-order accurate by construction. If our implementation of the Poisson solver is correct, therefore, the relative errors should be inversely proportional to the square of the grid spacing. To check if this is indeed the case, we repeat the uniform sphere tests by varying the number of cells from $16^3$ to $512^3$.
Figure \ref{fig:convergence} plots as circles the mean relative errors $\left\langle {\epsilon} \right\rangle$ from the sphere tests as functions of $N_z$. Overall, the errors decrease roughly at a second-order rate with increasing $N_z$, but exhibit some fluctuations.
\citet{katz16} noted that these fluctuations of the errors are caused not by the truncation errors of the finite-difference scheme but by inability of an adopted grid to perfectly resolve a spherical mass distribution. This is true even for a spherical grid when the sphere center offsets from the origin.

To delineate the truncation errors alone, it is thus necessary to design a solid figure whose shape is identical to the cell shape of an adopted grid.  In addition, the size and mass of the solid figure should be unchanged with varying resolution. For this purpose, we consider a uniform cube with density $\rho=1$, located at $x_1 \le x \le x_2$, $y_1 \le y \le y_2$, and $z_1 \le z \le z_2$
in a Cartesian grid. In a cylindrical grid, we consider a rectangular torus with density $\rho=1$, occupying the regions with $R_1 \le R \le R_2$, $\phi_1 \le \phi \le \phi_2$, and $z_1 \le z \le z_2$. Table \ref{tb:convergence_test} lists the parameters of the solid figures that we adopt: these values ensure that the mass distribution does not change with resolution.

We use our Poisson solver to calculate the gravitational potentials of the solid figures by varying resolution from $N_z=16$ to $512$, while keeping the domain sizes the same as in Section \ref{s:staticPot}. As the reference potential, we take Equation (20) of \citet{katz16} for the gravitational potential of a uniform cube. There is no algebraic expression for the potential of a rectangular torus, but \citet{hure14} provided a closed-form expression, in terms of line integrals with smooth integrands, in their Equation (29). We use the Romberg's method with a relative tolerance $10^{-10}$ to ensure that the numerical integrations are accurate enough to serve as a reference solution. The squares in Figure \ref{fig:convergence} plot the resulting mean relative errors $\left<{\epsilon}\right>$ for the cube and rectangular torus. Note that $\left<{\epsilon}\right>$ against $N_z$ follows almost a straight line with slope of $-1.993$, $-2.003$, and $-2.003$ in the Cartesian, uniform cylindrical, and logarithmic cylindrical grid, respectively, confirming that our implementation of the Poisson solver retains a second-order accuracy.

\begin{deluxetable}{ccccccc}
  \tablecaption{Parameters for the convergence test \label{tb:convergence_test}}
  \tablehead{
    \colhead{Coordinate system} & \colhead{$x_1$ or $R_1$} & \colhead{$x_2$ or $R_2$} & \colhead{$y_1$ or $\phi_1$} & \colhead{$y_2$ or $\phi_2$} & \colhead{$z_1$} & \colhead{$z_2$}
  }
  \startdata
  Cartesian & 0.0625 & 0.4375 & -0.375 & 0 & -0.125 & 0.25\\
  uniform cylindrical & 0.625 & 0.90625 & 0 & 1.570796327 & -0.0625 & 0.1875\\
  logarithmic cylindrical & 0.1 & 0.7498942093 & 0 & 1.570796327 & -0.0625 & 0.1875
  \enddata
\end{deluxetable}

\subsection{Performance Test}

To check the parallel performance of our implementation, we conduct a weak scaling test of our Poisson solver on the {\tt TigerCPU} linux cluster at Princeton University\footnote{\url{https://researchcomputing.princeton.edu/systems-and-services/available-systems/tiger}}. The {\tt TigerCPU} cluster consists of 408 nodes, with each node comprised of 40 $2.4\,{\rm GHz}$ intel Skylake processors. We measure the wall clock time using the {\tt MPI\_Wtime} function, after a call to {\tt MPI\_Barrier} to synchronize all processors. We run the job with {\tt SLURM --exclusive} option to make sure that other jobs do not interfere with ours.

For the weak scaling test, we divide the whole computational domain into $N_{\rm core}$ subdomains consisting of $64^3$ cells each. In the {\tt Athena++} terminology, the whole domain and subdomain are referred to as {\tt Mesh} and {\tt MeshBlock}, respectively. We assign each {\tt MeshBlock} to a single processor, while varying $N_{\rm core}$ from $1$ to $4096$. The corresponding size of {\tt Mesh} varies from $64^3$ to $1024^3$. In each run, we call our Poisson solver, together with the MHD solver for comparison, 100 times, and then measure the wall clock time per cycle $t_{\rm wall}$ for various steps. We repeat the calculations five times in order to avoid unusual runs due to stale nodes.

\begin{figure}[htb!]
 \plotone{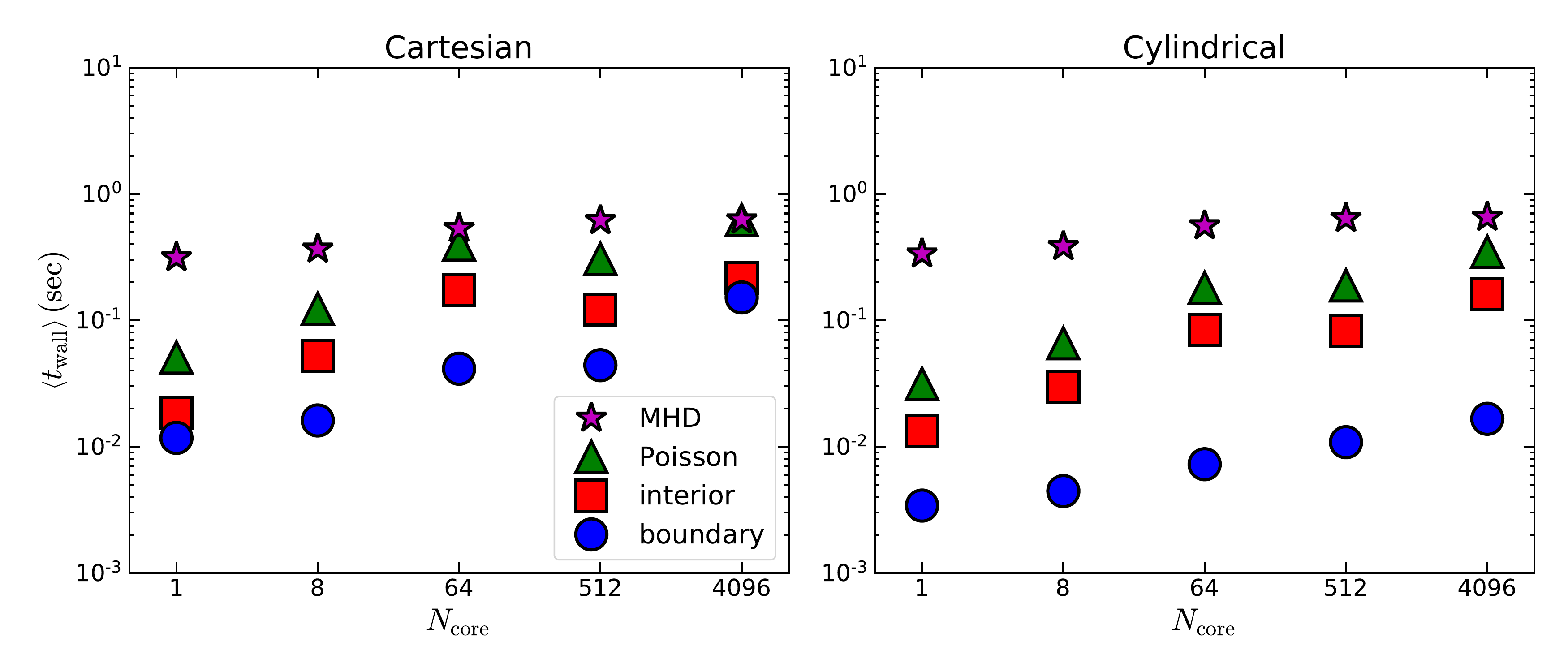}
  \caption{Average wall clock time per cycle \emph{vs.} the number of processors for runs with Cartesian (\emph{left}) and cylindrical (\emph{right}) grids. The symbols indicate the time taken by MHD solver (stars), Poisson solver (triangles), interior solver (squares), and the boundary solver (circles). Note that the time taken by the Poisson solver is equal to twice the time taken by the interior solver plus the time taken by the boundary solver.}
  \label{fig:scaling}
\end{figure}

Figure \ref{fig:scaling} plots the mean values $\left\langle t_{\rm wall} \right\rangle$ of the wall clock times per cycle as functions of $N_{\rm core}$ for the Cartesian (left) and cylindrical (right) grids.  As noted earlier, our Poisson solver requires to run the interior solver twice and the boundary solver once. The total time taken by the Poisson solver (triangles) is dominated by the interior solver (squares) rather than the boundary solver (circles). While the time taken by the MHD solver is comparable between Cartesian and cylindrical grids, the Poisson solver is more efficient in the cylindrical grid. This is because Cartesian coordinates has no inherent periodic direction and thus requires more operations to implement the open boundary conditions.  In addition, the Cartesian grid has two more
boundaries than the cylindrical grid and thus needs more
boundary-to-boundary interactions in the boundary solver.
Notwithstanding these differences,
the Poisson solver in both Cartesian and cylindrical grids takes less time than the MHD solver (stars) at least up to 4096 processors, leading us to conclude that our Poisson solver is very efficient and does not contribute much to the total computational cost of self-gravitating MHD simulations.\footnote{The weak scaling test shown in Figure \ref{fig:scaling} hints some performance degradation from $N_{\rm core}=1$ to $64$ relative to $\langle t_{\rm wall}\rangle \propto \ln N_{\rm core}$ expected for the theoretical FFT. Our parallel FFT utilizes a ``transpose algorithm'' known to be efficient when a data size for communication is larger than the critical size that depends
on the latency/bandwidth of the interconnecting network device and
the network topology \citep[e.g.,][]{foster97}.  An alternative ``binary exchange algorithm'' may work efficiently for a small data size \citep[e.g.,][]{muller19}.}

\subsection{Ring Fragmentation}\label{s:simulation}

\begin{figure}[ht]
  \plotone{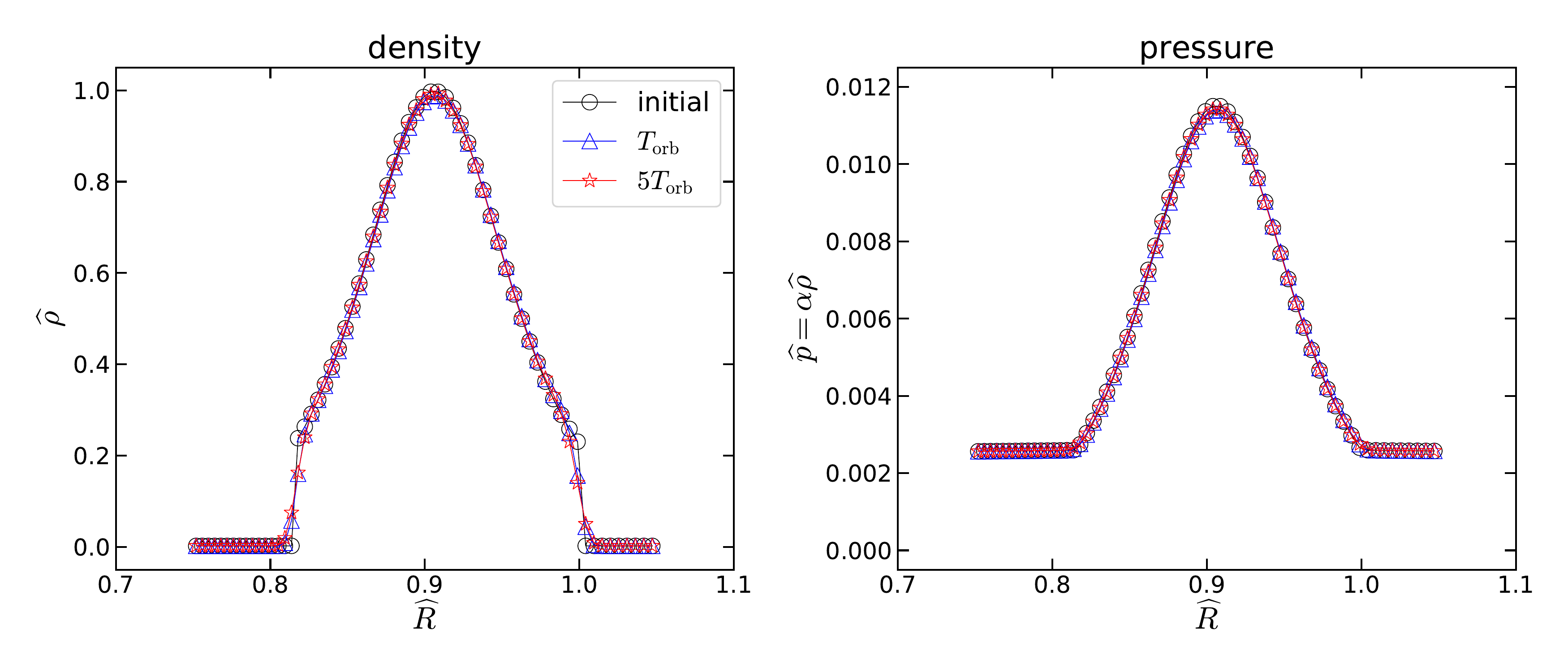}
  \caption{Density and pressure profiles of the ring at initial time (circles) and after one (triangles) and five (stars) orbital periods. Except for small changes near the boundaries, the ring without any perturbations remains almost intact.}
  \label{fig:torus_ic}
\end{figure}

As a final test of our Poisson solver, we run time-dependent simulations for
fragmentation of a self-gravitating isothermal ring using {\tt Athena++}.
As an initial condition, we consider a rigidly-rotating ring at angular velocity
$\Omega_0$ and sound speed $c_s$, surrounded by a tenuous hot external medium. The ring is initially in hydrosteady equilibrium under both self-gravity and external gravity. Assuming that the external gravity ${\bf g}_{\rm ext} = - \Omega_e^2 {\bf R}$ alone makes the ring rotate at angular frequency $\Omega_e$,  the equation for such an equilibrium reads
\begin{equation}\label{eq:hydrosteady}
  c_s^2 \nabla \ln \rho + \nabla \Phi - \Omega_s^2{\bf R} = 0,
\end{equation}
together with Equation \eqref{eq:Poisson}, where $\Omega_s \equiv (\Omega_0^2 - \Omega_e^2)^{1/2}$ is the angular velocity due to self-gravity alone \citep{kim16}. One can show that
the equilibrium configurations are completely specified by two
dimensionless parameters: $\alpha\equiv c_s^2/(GR_A^2\rho_c)$
and $\widehat{\Omega}_s \equiv \Omega_s/(G\rho_c)^{1/2}$, where
$\rho_c$ and $R_A$ denote the
maximum density and the maximum radial extent of
an equilibrium object, respectively.

Using the self-consistent field method of \citet{hachisu86},
we solve Equations \eqref{eq:Poisson} and \eqref{eq:hydrosteady}
alternatively and iteratively to find the equilibrium configuration
for $\alpha=0.015$ and $\widehat{\Omega}_s=0.22$. We then boost the angular velocity of the ring to $\widehat{\Omega}_0=0.3$ to account for ${\bf g}_{\rm ext}$.
The external medium is set to be hotter than the ring by two orders of magnitude.
In order to check whether the configuration we set is really in equilibrium,
we evolve it over time on
a logarithmic cylindrical grid with size $\widehat{R}\equiv R/R_A \in[0.75,1.05]$, $\phi\in[0,2\pi]$,
and $z/R_A \in[-0.15,0.15]$,  without imposing any perturbations.
The number of cells used is $N_R=64$, $N_\phi = 1024$, and $N_z=64$.
Figure \ref{fig:torus_ic} compares
the radial profiles of the dimensionless
density $\widehat{\rho}\equiv\rho/\rho_c$ and
pressure $\widehat{p}=\alpha\widehat{\rho}$ in
the equatorial plane as functions of
$\widehat{R}$, at $t/T_{\rm orb}=0$, 1, and 5,
where $T_{\rm orb}=2\pi/\Omega_0$.
The density and pressure profiles are almost unchanged over time
except near the contact discontinuity between the ring and external medium,
demonstrating that the initial steady configuration is well maintained over many orbital periods.

\begin{figure}[ht]
 \plotone{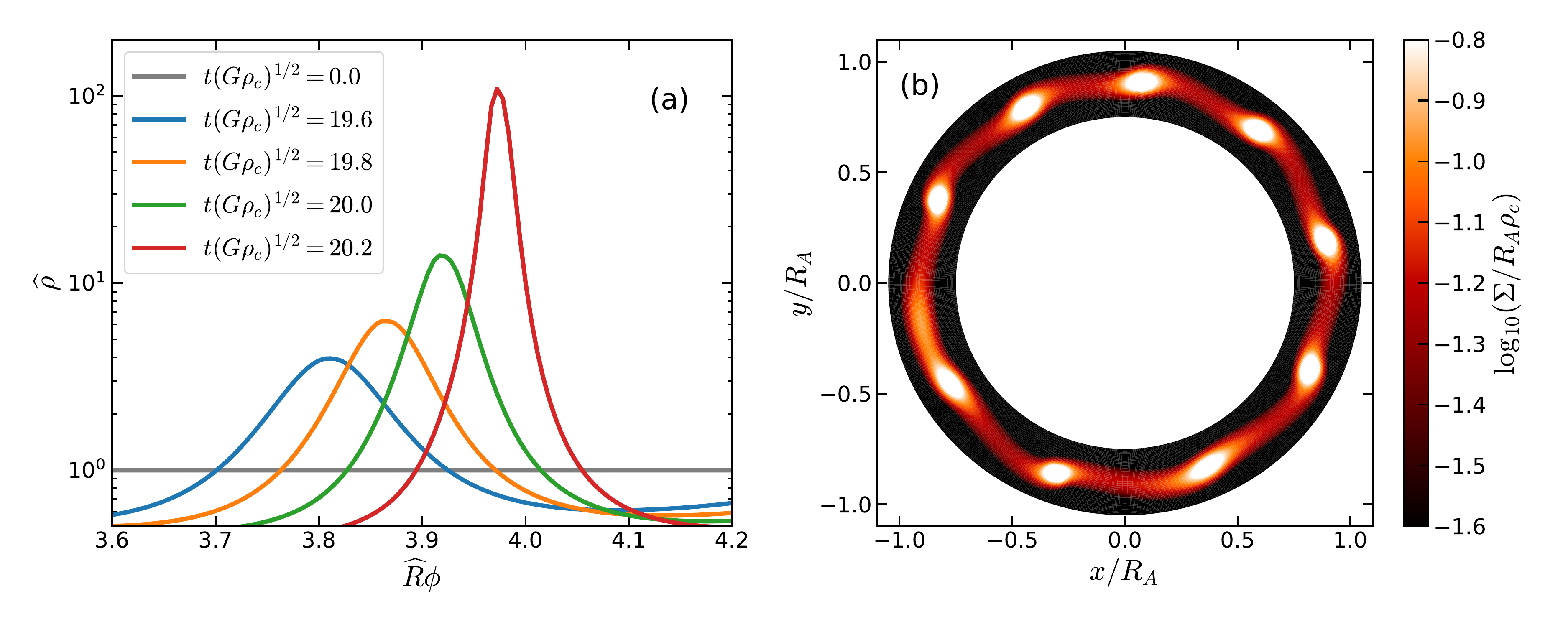}
  \caption{(a) Temporal changes of the azimuthal density profile at $\widehat{R} = 0.91$ and $z=0.0023$. (b) Projected surface density $\Sigma=\int \rho\, dz$ at time $t(G\rho_c)^{1/2} = 20.2$.}
  \label{fig:gi}
\end{figure}

By performing a linear-stability analysis,
\citet{kim16} found that the above equilibrium is gravitationally unstable to non-axisymmetric perturbations for a range of the azimuthal mode number $m$. The most unstable modes were found to have $m=9$ and $10$, with an almost equal growth rate ${\rm Im}(\omega)\approx 0.8(G\rho_c)^{1/2}\approx 17\,T_{\rm orb}^{-1}$
and a phase speed ${\rm Re}(\omega)/m\approx 0.3(G\rho_c)^{1/2}$, corresponding to overstability. To check if our Poisson solver can
pick up these unstable modes and capture their growth, we generate
random density perturbations with amplitude $10^{-5}$
and apply them to the equilibrium density $\rho$.
We then evolve the system and monitor how various modes grow.
Figure \ref{fig:gi} plots the azimuthal profiles of the dimensionless
density at $\widehat{R}=0.91$ and $\widehat{z}=0$ at a few selected epochs
as well as the surface density $\Sigma=\int \rho\,dz$ at $t(G\rho_c)^{1/2}=20.2$.
It is apparent that the perturbations grow as they propagate along the $\phi$-direction. In the highly nonlinear stage, the density distribution is dominated
equally by the $m=9$ and $m=10$ modes, indicating that
these are two fastest growing modes of the instability.

\begin{figure}[ht]
 \plotone{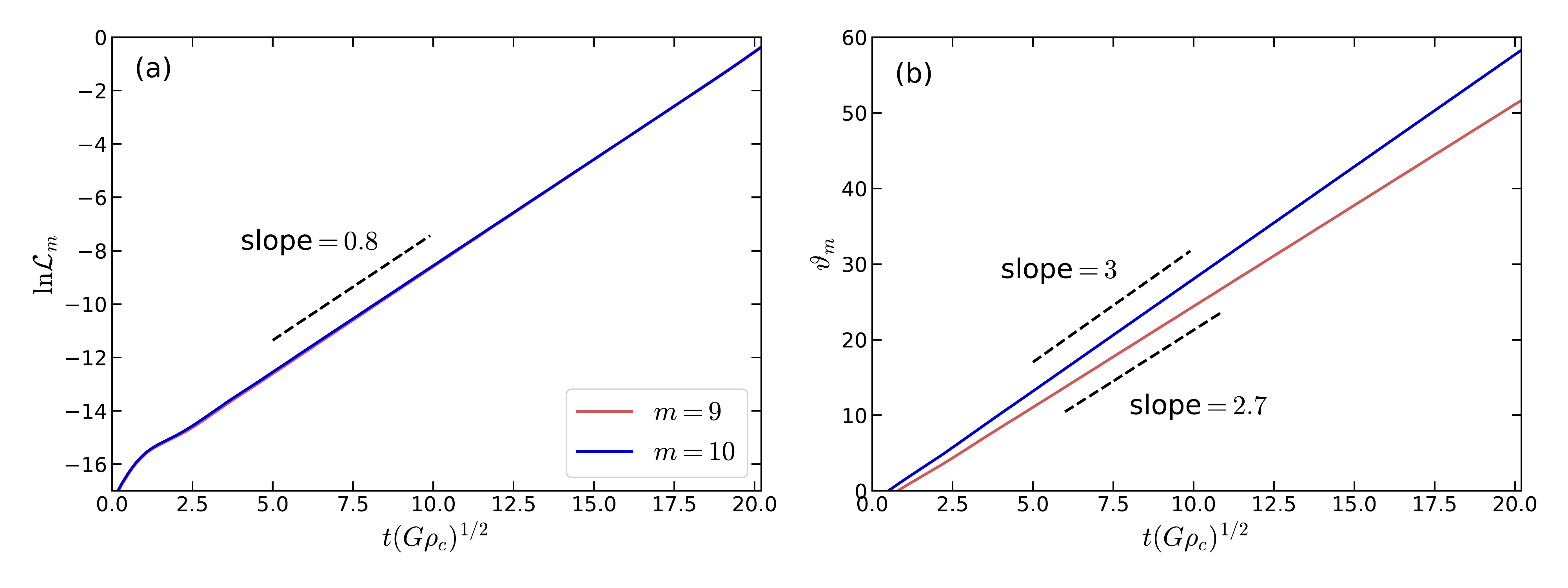}
  \caption{Time evolution of the Fourier amplitude ${\cal L}_{m}$ and the phase angle $\vartheta_m$ for the $m=9$ and $10$ modes. The dashed lines correspond to a slope
of 0.8 for the linear growth of both ${\cal L}_{9}$ and ${\cal L}_{10}$ in (a), and
2.7 and 3.0 for the constant changes of $\vartheta_9$ and $\vartheta_{10}$, respectively, in (b), consistent with the result of linear stability analysis of \citet{kim16}.
  }
  \label{fig:growth}
\end{figure}

To measure the growth rate and the phase speed of each mode in the numerical simulation, we calculate the Fourier transform  ${\cal L}_me^{-i\vartheta_m} \equiv \int \bar{\rho}(\phi) e^{-im\phi}\,d\phi  $ of the radially- and vertically-integrated density $\bar{\rho}(\phi) = \iint\rho\,dR dz$. Figure \ref{fig:growth} plots the temporal variations of the Fourier amplitude ${\cal L}_{m}$ and the phase angle $\vartheta_{m}$ for $m=9$ and $m=10$.  The two modes grow exponentially as they propagate at a constant phase speed. The dimensionless growth rate $d \ln {\cal L}_m/dt/(G\rho_c)^{1/2}$ and the phase speed $d {\vartheta_m}/dt/[m(G\rho_c)^{1/2}]$ are measured to be $0.80$ and $0.30$, respectively, consistent with the results of the linear stability analysis.

Taken together, all the test results presented in this section
demonstrate that our 3D cylindrical Poisson solver is reliable, accurate to second order, and very efficient.

\section{SUMMARY AND DISCUSSION}\label{s:discussion}

To study dynamical evolution of self-gravitating, rotating disks, it is desirable to use a fully 3D Poisson solver in a cylindrical geometry subject to the open boundary condition. In this paper, we have presented an accurate and efficient algorithm for
such a Poisson solver that works in Cartesian, uniform cylindrical, and logarithmic cylindrical coordinates.
Our algorithm adopts ``surface screening charge'' method introduced by
James, employing the DGF
to calculate the boundary potential consistent with
the open boundary condition (Section \ref{s:bc}), and utilizes the
eigenfunction expansion method to solve the interior potential (Section \ref{s:interior_solver}).
The computational cost of our algorithm is of order ${\cal O}(N^3 + N^2\log N)$, with $N$ being the number of cells in one dimension.
The results of the various test problems presented in Section \ref{s:tests} confirm that our Poisson solver is second-order accurate and takes
less computational cost than the MHD solver in {\tt Athena++} up to 4096 cores.

\begin{figure}[ht]
  \epsscale{0.6}
 \plotone{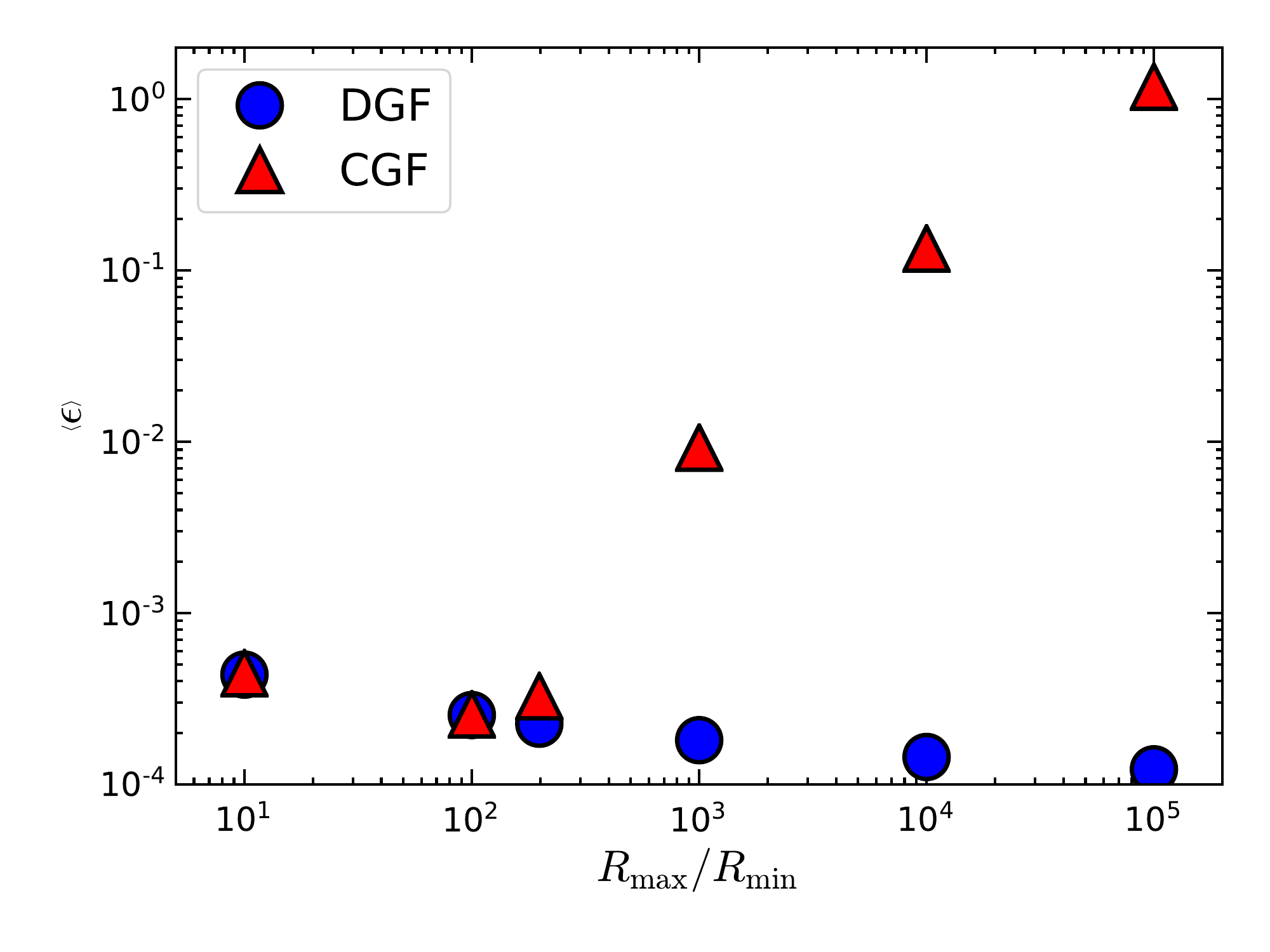}
  \caption{Mean relative errors of the gravitational potential based on the CGF (red triangle) and the DGF (blue circles) for a rectangular torus as functions of  $R_{\rm max}/R_{\rm min}$. The computational domain and the torus parameters are kept the same as given in Tables \ref{tb:sphere_test} and \ref{tb:convergence_test}, although $R_{\rm min}$ is
changed to vary $R_{\rm max}/R_{\rm min}$ from $10$ to $10^5$.
  }
  \label{fig:Rmaxmin}
\end{figure}

The second-order accuracy of our Poisson solver is enabled by the usage of the cylindrical DGF.
Although \citet{serafini05} and \citet{snytnikov11} implemented the James algorithm in Cartesian and cylindrical coordinates, respectively, they simply used the CGF instead of the DGF after enlarging the computational domain. Since the difference between CGF and DGF is quite small at large distances from the source (see Appendix \ref{s:calc_dgf}), the potential based on the CGF would be similar to that with the DGF if the domain is expanded sufficiently. As mentioned in Introduction, however, domain expansion in cylindrical coordinates is very limited toward the inner radial boundary, so that the potential with the CGF would become less accurate for larger $R_{\rm max}/R_{\rm min}$, where $R_{\rm max}$ and $R_{\rm min}$ denote the outer and inner radial boundaries of the domain, respectively. To demonstrate this, we recalculate the gravitational potential of a rectangular torus on a logarithmic cylindrical grid presented in Section \ref{s:convergence}, but this time by using the CGF. For fare comparison, we apply domain expansion technique similar to \citet{snytnikov11} when using CGF. Specifically, we use the same extended domain for DGF calculation defined in Appendix \ref{s:calc_dgf_cyl} for the enlarged domain. Figures \ref{fig:Rmaxmin} plots the resulting mean relative errors as red triangles against $R_{\rm max}/R_{\rm min}$, in comparison with the cases based on the DGF plotted as blue circles. The CGF works as well as the DGF for $R_{\rm max}/R_{\rm min} \lesssim 10^2$. But, it fails to give second-order convergence for systems with $R_{\rm max}/R_{\rm min} \gtrsim 10^2$, which are common in astronomical applications \citep[e.g.,][]{kuiper10,seo13,zhu12,bae14,ju16,kim17}.  In these cases,
it is necessary to use the cylindrical DGF for accurate potential calculations.

\acknowledgments
SM wishes to thank Chang-Goo Kim, Jeong-Gyu Kim and Kengo Tomida for their helpful discussions and advice. The work of SM was supported by NRF (National Research Foundation of Korea) Grant funded by the Korean Government (NRF-2017-Fostering Core Leaders of the Future Basic Science Program/Global Ph.D. Fellowship Program).
The work of WTK was supported by the grant (2017R1A4A1015178) of National Research Foundation of Korea. The work of ECO on this project is supported by grant 510940 from the Simons Foundation. The computation of this work was supported by the Supercomputing Center/Korea Institute of Science and Technology Information with supercomputing resources including technical support (KSC-2018-C3-0015) and  the PICSciE TIGRESS High Performance Computing Center at Princeton University.

\software{{\tt Athena++} (J. M. Stone et al. 2019, in preparation, \url{http://princetonuniversity.github.io/athena/}), {\tt IPython} \citep{perez07}, {\tt FFTW} (\url{www.fftw.org}), Plimpton's parallel transpose (\url{https://www.sandia.gov/~sjplimp/docs/fft/README.html})}

\appendix

\section{SECOND-ORDER FINITE-DIFFERENCE IN LOGARITHMIC CYLINDRICAL COORDINATES}\label{s:fd}

Here we derive a second-order finite-difference approximation to the radial part of the Laplace operator in logarithmic cylindrical coordinates. With the change of the variables $u\equiv \ln R$, the radial part of the Laplacian becomes
\begin{equation}
  \frac{1}{R} \frac{\partial}{\partial R} \left( R \frac{\partial\Phi}{\partial R}  \right) = \frac{1}{R^2} \frac{\partial^2\Phi}{\partial u^2}.
\end{equation}
Since the logarithmic grid in $R$ ($R_i = R_0 f^i$) corresponds to a uniform grid in $u$ ($u_i = u_0 + i\ln f$), we can apply a centered difference scheme in the $u$-space to obtain
\begin{equation}\label{eq:radiald_alt}
  \frac{1}{R^2} \frac{\partial^2\Phi}{\partial u^2} = \frac{1}{R_i^2} \frac{\Phi_{i-1} - 2\Phi_i + \Phi_{i+1}}{(\delta u)^2} + {\cal O}((\delta u)^2).
\end{equation}
Noting that $\delta u = \ln f = N_R^{-1}\ln (R_{\rm max}/R_{\rm min})$, the above expression can be expressed in the $R$-space as
\begin{equation}\label{eq:radiald2}
  \frac{1}{R} \frac{\partial}{\partial R} \left( R \frac{\partial\Phi}{\partial R}  \right) = \frac{\Phi_{i-1}-2\Phi_i + \Phi_{i+1}}{(R_i\ln f)^2} + {\cal O}\left( \frac{1}{N_R^2} \left(\ln\frac{R_{\rm max}}{R_{\rm min}}\right)^2   \right).
\end{equation}
It is evident that the remainder decreases at a second-order rate with increasing $N_R$. We adopt Equation \eqref{eq:radiald2} as our discrete Laplace operator in the logarithmic cylindrical grid (see Equation \eqref{eq:radial_operator}).

\section{COMPUTATION OF THE DISCRETE GREEN'S FUNCTION}\label{s:calc_dgf}

The DGF is needed in order to calculate the surface potential associated with
the surface screening charges.  The DGF is pre-computed once at the
beginning of any simulation.  In this Appendix we outline the numerical
method used to evaluate the DGF.

\subsection{Cartesian Grid}\label{s:calc_dgf_cart}

We calculate the DGF ${\cal G}_{i-i',j-j',k-k'}$ in Cartesian coordinates by solving Equation \eqref{eq:def_car_green} for a point source with unit mass located in a cell at $(i',j',k')=(1,1,1)$.
Once we obtain ${\cal G}_{i-1,j-1,k-1}$  for $1\le i \le N_x$,
$1\le j \le N_y$, and $1\le k \le N_z$, the values for other indices can be computed from the symmetry requirement
\begin{equation}\label{eq:car_symm}
{\cal G}_{i-i',j-j',k-k'} = {\cal G}_{|i-i'|,|j-j'|,|k-k'|},
\end{equation}
for $| i-i'| \le N_x-1$, $|j-j' | \le N_y-1$, and $ | k-k' | \le N_z-1$.

As we use the method presented in Section \ref{s:interior_solver} to
obtain the numerical solution of Equation \eqref{eq:def_car_green},
we must supply an appropriate boundary condition a priori.

It is reasonable to assume that far from the source, the DGF asymptotes to the CGF:
\begin{equation}\label{eq:car_asymptotic_green}
  {\cal G}_{i-1,j-1,k-1} \approx -\frac{G}{\sqrt{(x_i - x_{1})^2 + (y_j - y_{1})^2 + (z_k - z_{1})^2 }}\quad\text{(far from the source).}
\end{equation}
Since high-order derivatives of the CGF are non-negligible close to the source, the near-field DGF that results from a second-order finite-difference approximation to the Poisson equation would deviate greatly from the CGF. How far should the boundary be away from the source to safely apply Equation \eqref{eq:car_asymptotic_green} as a proper boundary condition? The answer of \citet{james77} to this question was 16 cells. In fact, one can compare the first two terms in the asymptotic expansion of the DGF in \citet{burk97} to verify that the  relative deviation between the DGF and CGF is $0.1\%$ at a distance of $16$ cells, regardless of the total number of cells or the grid spacing.

To ensure that all boundaries are sufficiently away from the point source at $(i',j',k') = (1,1,1)$, for the calculation of the DGF only, we extend the computational domain by adding 16 additional cells to one side in each direction. The newly added cells
(except for the ghost cells) are numbered as $i=-15,-14,\cdots, -1$ in the $x$-direction (and similarly in the $y$- and $z$-directions). We then apply the boundary condition (Equation \eqref{eq:car_asymptotic_green}) to $i=-15$ and $N_x+1$ (and similarly for $j$ and $k$) and solve Equation \eqref{eq:def_car_green} in the extended domain to obtain ${\cal G}_{i-1,j-1,k-1}$ for $i=-15,-14,\cdots,N_x+1$ (and similarly for $j$ and $k$).  In practice, following the method of
Section \ref{s:interior_solver} we use the boundary
condition to define a modified interior charge, and then employ sine
transforms.

\begin{figure}[ht]
 \plotone{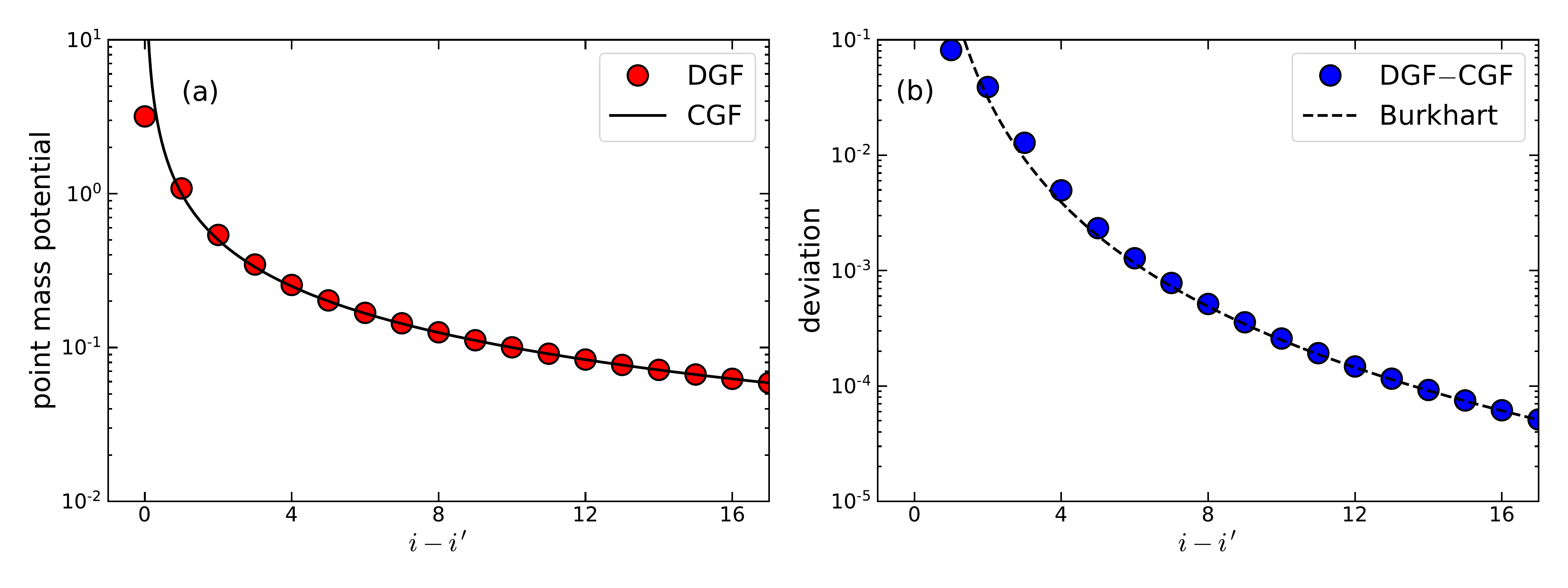}
  \caption{Comparison between the DGF and CGF, normalized by $-G/\delta x$,
 in Cartesian coordinates. (a)
DGF (red circles) and CGF (black solid line)
as functions of the grid distance from the source located at $i=0$. (b) The difference between the DGF and CGF (blue circles) and the Burkart's asymptotic expansion (black dashed line; the term involving $\eta_3$ in their Equation (7), where we take $K=G$ to be consistent with our definition of the Green's function).
  }
  \label{fig:cardgf}
\end{figure}

After obtaining the solution, we discard the portion pertaining to the extended part of the domain, and use Equation \eqref{eq:car_symm} to calculate the Cartesian DGF for whole indices. Since the DGF is calculated once and for all in the initialization step, its contribution to the computational cost of an entire simulation is almost negligible. Figure \ref{fig:cardgf} plots the resulting DGF as a function of the grid distance from the source, in
comparison with the CGF. The CGF diverges at the source position, whereas the DGF remains finite everywhere. Although the DGF deviates from the CGF close to the source,
their difference becomes smaller as the grid distance increases, consistent with
the asymptotic expansion of \citet{burk97}.

\subsection{Cylindrical Grid}\label{s:calc_dgf_cyl}

In a cylindrical grid, we need to store only 16 kinds of the DGFs, such as ${\cal G}_{i,k'}^m (\rm inn \rightarrow top)$, appearing in Equation \eqref{eq:Theta_top}--\eqref{eq:Theta_out}, each of which represents the Fourier-transformed potential at one boundary due to point masses located at the same or different boundaries. These are calculated as follows.

To calculate the potential generated by point sources at the top vertical boundary,
we place a point source with unit mass at
$(i',j',k') = (i',1,N_z+1)$ for any $i' \in [1, N_R]$ and
solve Equation \eqref{eq:def_cyl_green} numerically using the
method of Section \ref{s:interior_solver}.
This yields ${\cal G}_{i,i',j-1,k-N_z-1}$ and its Fourier
transform ${\cal G}_{i,i',k-N_z-1}^m$ after applying Equation \eqref{eq:dft},
but we only keep four 3D arrays
${\cal G}^m_{i,i'}({\rm top \to top}) = {\cal G}^m_{i,i',0}$,
${\cal G}^m_{i,i'}({\rm top \to bot}) = {\cal G}^m_{i,i',-N_z-1}$,
${\cal G}^m_{k,i'}({\rm top \to inn}) = {\cal G}^m_{0,i',k-N_z-1}$, and
${\cal G}^m_{k,i'}({\rm top \to out}) = {\cal G}^m_{N_R+1,i',k-N_z-1}$,
corresponding to four DGFs due to the point sources at the top boundary. We repeat the above calculations by varying $i' \in [1, N_R]$ to fill all the components of the four DGFs.

For the potential generated by point sources at the inner radial boundary,
we place a point mass at $(i',j',k')=(0,1,k')$ for any $k' \in [1, N_z]$.
We then solve Equation \eqref{eq:def_cyl_green} and apply
Fourier transform to obtain
${\cal G}_{i,0,k-k'}^m$.  We keep only four 3D arrays
${\cal G}^m_{i,k'}({\rm inn\to top}) = {\cal G}^m_{i,0,N_z+1-k'}$,
${\cal G}^m_{i,k'}({\rm inn\to bot}) = {\cal G}^m_{i,0,-k'}$,
${\cal G}^m_{k-k'}({\rm inn\to inn}) = {\cal G}^m_{0,0,k-k'}$, and
${\cal G}^m_{k-k'}({\rm inn\to out}) = {\cal G}^m_{N_R+1,0,k-k'}$\footnote{Although ${\cal G}^m_{k-k'}({\rm inn\to inn})$ and ${\cal G}^m_{k-k'}({\rm inn\to out})$ can be stored in 2D arrays, we store them as 3D arrays for simple coding.},
corresponding to four DGFs due to the point sources at the inner radial boundary.
We repeat the above calculations for all $k'\in[1,N_z]$ to
completely fill the elements of the arrays.
We follow a similar procedure to obtain
the remaining eight DGFs due to points sources at the bottom
and outer radial boundaries.

The boundary condition in solving Equation \eqref{eq:def_cyl_green}
can be obtained by requiring that the DGF at a large distance from the source
is approximately equal to
\begin{equation}\label{eq:cyl_asymptotic_green}
  {\cal G}_{i,i',j-j',k-k'} \approx - \sum_{p=0}^{P-1} \frac{G}{\sqrt{R_i^2 + R_{i'}^2 - 2R_iR_{i'}\cos(\phi_j - \phi_{j'} - pL_\phi) + (z_k - z_{k'})^2 }}\quad\text{(far from the source)},
\end{equation}
The summation over $p$ in Equation \eqref{eq:cyl_asymptotic_green} is to add all the contributions from the periodic images of the point mass when the mass distribution holds $P$-fold symmetry in the $\phi$-direction.

As discussed in Section \ref{s:calc_dgf_cart}, Equation \eqref{eq:cyl_asymptotic_green} remains valid as long as the distance between the cells $(i,j,k)$ and $(i',j',k')$ is sufficiently large. Similarly to the Cartesian case, we extend the computational domain, only for the calculation of the DGF, by adding extra cells to each of the four boundaries. In the vertical and outer radial directions, 16 cells are wide enough to ensure that Equation \eqref{eq:cyl_asymptotic_green} is a good approximation to the DGF. In the inner radial direction, however, the 16-cell criterion based on a uniform grid spacing does not guarantee that Equation \eqref{eq:cyl_asymptotic_green} is a valid approximation especially when $f=(R_{\rm max}/R_{\rm min})^{1/N_R}$ is large. We numerically confirmed that using Equation \eqref{eq:cyl_asymptotic_green} as a Dirichlet boundary condition at the extended inner radial boundary causes large errors in the computation of the cylindrical DGF and ultimately gravitational potential returned from our Poisson solver when $R_{\rm max}/R_{\rm min} > 10^2$.

\begin{figure}[htb!]
 \plotone{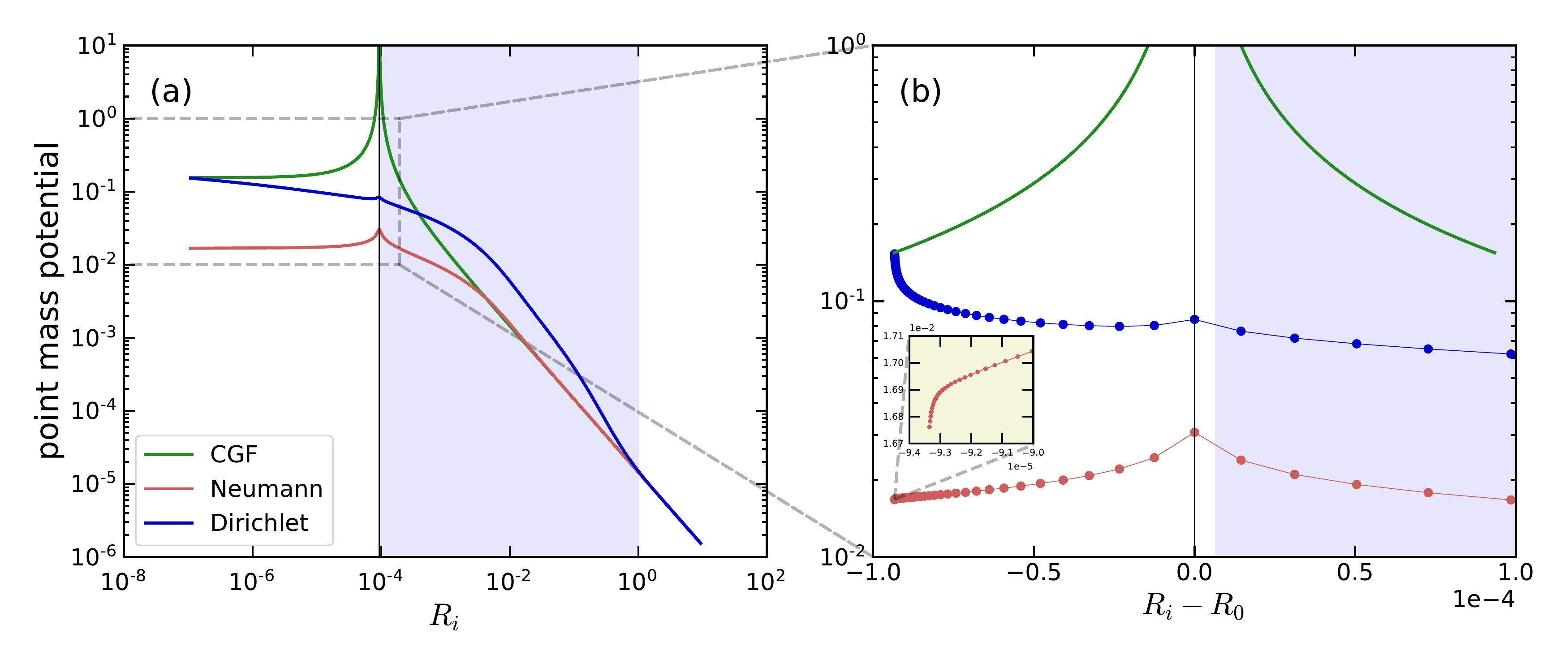}
  \caption{DGF due to a point mass at $(i',j',k') = (0,1,1)$ with the abscissa in the (a) logarithmic and (b) linear scale. The original computational domain, indicated by shade, is a logarithmic cylindrical grid of $64^3$ cells spanning $R\in[10^{-4},1]$, $\phi\in[0,2\pi]$, $z\in[-0.25,0.25]$. The black vertical line marks the position of the point mass, which is the innermost ghost cell of the original domain. The green solid lines draw the CGF from which the boundary condition is obtained. Red and blue lines are the DGF resulting from the Neumann and the Dirichlet boundary condition, respectively. Panel (b) zooms in the regions with $|R_i-R_{-1}| \le 10^{-4}$ in (a). The inset in (b) shows the zoom-in view near the extended inner boundary (with the ordinate in linear scale). All three Green's functions are dimensionless, normalized by $-G/\delta R_{0}$.
}
  \label{fig:dgf}
\end{figure}

It turns out that using the radial gradient of Equation \eqref{eq:cyl_asymptotic_green} as a Neumann boundary condition is a cure to this inner boundary problem.
The Neumann condition on the extended inner boundary is not very accurate, either, but we empirically find that it enables the DGF to converge rapidly to the desired values and results in the accurate DGF at least in the original computational domain, if the domain is sufficiently expanded. We found that adding $N_R-16$ cells in the inner radial boundary ($16$ cells are kept for the outer radial boundary) produces enough accuracy for the DGF in the original domain.
To illustrate this directly,
we set up a logarithmic cylindrical grid with $64^3$ cells spanning $R\in[10^{-4},1]$,
$\phi\in[0,2\pi]$, $z\in[-0.25,0.25]$.
We then distribute additional $N_R=64$ cells, by adding 16 cells with $i=65, 66,\cdots,80$ to outside of the outer radial boundary
and $N_R-16 = 48$ cells with $i=-47, \cdots,-1,0$ to inside of the inner radial boundary,
and calculate the DGF for a point source located at $(i',j',k')=(0,1,1)$
for both Neumann and Dirichlet conditions at the extended inner boundary.
Figure \ref{fig:dgf}(a) compares the resulting ${\cal G}_{i,0,0,0}$ as functions of $R_i$ in logarithmic scale,
with crosses and squares corresponding to the cases with Neumann and Dirichlet conditions, respectively.
Figure \ref{fig:dgf}(b) zooms in the section with $|R_i-R_{-1}| \le 10^{-4}$ into linear scale,
with the inset plotting the DGF
from the Neumann condition
for $-9.4\times 10^{-5} \le R_i - R_{-1} \le -9.0\times 10^{-5}$.
It is clear that the DGF resulting from the Dirichlet condition is
not symmetric with respect to the source, and cannot thus be considered correct.
On the other hand, the DGF from the Neumann condition retains symmetry with
respect to the source and converges to the CGF at large distances toward the outer boundary.
The test problems presented in Section \ref{s:tests} verify that
the DGF under the Neumann condition works very well for our purposes,
yielding accurate gravitational potentials in the original domain even when $R_{\rm max}/R_{\rm min} = 10^{5}$ (see Figure \ref{fig:Rmaxmin}).

In the case of a uniform cylindrical grid, it is sufficient to add only $16$ cells to inside of the inner radial boundary either using Neumann or Dirichlet boundary condition\footnote{We use Neumann condition for consistency.}. This is because $16$ cells are wide enough at both inner and outer boundary with uniform spacing. Since the radial coordinate of the extended inner boundary should be positive, this requires $R_{\rm min} > 16\delta R$, or equivalently, $R_{\rm max}/R_{\rm min} < 1+N_R/16$. This limitation is not be severe, given that the logarithmic cylindrical grid is more appropriate for large $R_{\rm max}/R_{\rm min}$.

\section{For Mass Distribution Under $P$-fold Azimuthal Symmetry}\label{s:P-fold_symm}

We often meet a problem that is periodic in the $\phi$-direction, with period $2\pi / P$ for a positive
integer $P$. For example, gas flows in barred galaxies may have $P=2$ symmetry (e.g., \citealt{seo13}),
while $P=4$ for four-armed spiral galaxies \citep[e.g.,][]{dobbs06,so08}.
In such situations, one can save computational time by restricting the domain to $\phi\in[0,2\pi/P]$, with a periodic boundary condition.

The interior solver presented in Section \ref{s:interior_solver_cylindrical} works without modification even when $L_\phi=2\pi/P$ does not cover a full $2\pi$ domain, because the $P$-fold symmetry is automatically taken into account in Equation \eqref{eq:cyl_eigenfunction}. This holds true also for the boundary solver (Section \ref{s:bc}) as long as the periodic boundary condition with period $L_\phi$ is imposed in solving Equation \eqref{eq:def_cyl_green}. The cylindrical DGF defined as such then represents the gravitational potential from $P$ identical point masses lying along the azimuth with a uniform angular separation of $L_\phi$. When applying the boundary condition to the cylindrical DGF, therefore, one should add all the contributions from the image masses located at $\phi\in[L_\phi, 2\pi]$ to the gravitational potential of a real point mass located in the original domain with $\phi\in[0, L_\phi]$, as in Equation \eqref{eq:cyl_asymptotic_green}.
We note that the inclusion of the image masses is implicit in the DGF, so that all calculations are done in the original computational domain with size $L_\phi$, enabling a factor of $P$ reduction in the computational time as well as memory compared to the cases with a full $2\pi$-periodic domain.

\begin{figure}[ht]
 \plotone{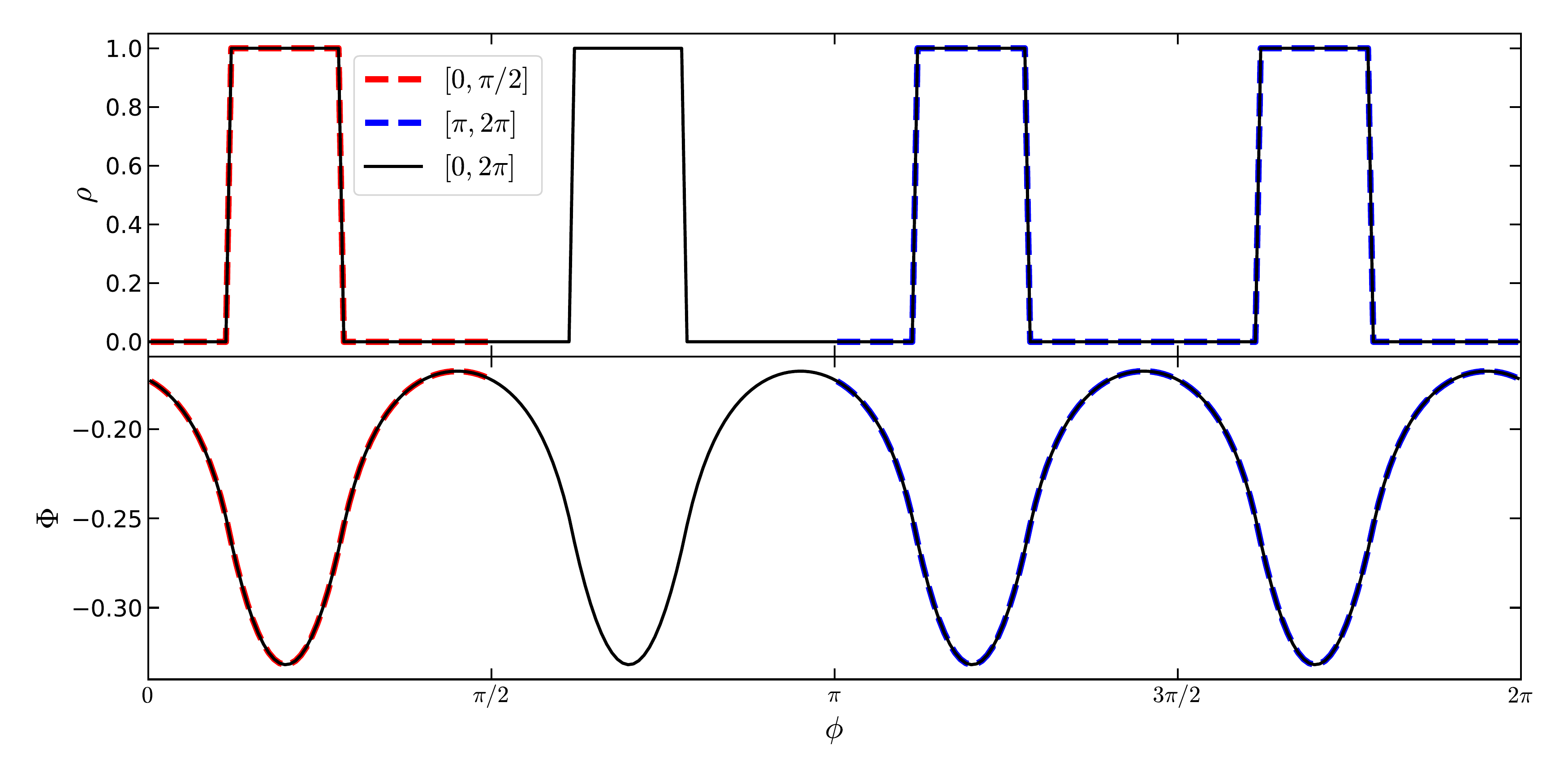}
  \caption{Azimuthal profiles of the mass density and the gravitational potential of four equally-spaced uniform spheres in $\phi$. The black solid lines are the potential-density pair at $R = 0.75$ and $z = 0.0039$ computed in the full $2\pi$ domain, while the red and blue dashed lines are those from the restricted domains with $\phi\in[0, \pi/2]$ and $[\pi, 2\pi]$, respectively. The three lines in the respective $\phi$-range agree with each other within the machine precision.
  }
  \label{fig:pfold}
\end{figure}

To verify that our method handles $P$-fold azimuthal symmetry, we consider four uniform spheres located at $\phi_0$, $\phi_0+\pi/2$, $\phi_0+\pi$, and $\phi_0+3\pi/2$ in a uniform
cylindrical grid presented in Table \ref{tb:sphere_test}. We calculate the gravitational potential from the resulting mass distribution that clearly has $P=4$ symmetry over $\phi\in[0, 2\pi]$. Figure \ref{fig:pfold} plots as black solid lines the mass density and the gravitational potential along the $\phi$-direction at $R=0.75$ and $z=0.0039$. We then recompute the gravitational potential of a single sphere at $\phi_0$ by reducing the computational domain to $\phi\in[0, \pi/2]$, and finally the gravitational potential of two spheres at $\phi_0+\pi$ and $\phi_0+3\pi/2$ in the domain covering $\phi\in[\pi, 2\pi]$.
The resulting potentials, plotted as red and blue dashed lines, are identical within machine precision to the potential from the full $2\pi$ domain. This confirms that our Poisson solver in a restricted $\phi\in[0,2\pi/P]$ domain correctly deals with mass distributions under $P$-fold azimuthal symmetry.

\bibliography{mybib}

\end{document}